\documentclass[aps,prl,twocolumn,secnumarabic,notitlepage,superscriptaddress,nofootinbib,tightenlines,floatfix]{revtex4-1}

\usepackage{comment}
\usepackage[pdftex]{color}
\usepackage{graphicx,capt-of}
\usepackage{amsmath,amssymb,amsfonts,mathtools}
\usepackage[colorlinks=true,linkcolor=black,filecolor=black,urlcolor=blue,citecolor=blue,pdftex,plainpages=false]{hyperref}

\def\Zmatch{Z_{\mathrm{match}}}
\newcommand{\Gnorm}{G^\mathrm{norm}}
\newcommand{\tauf}{\tau_\mathrm{F}}
\def\MSbar{\overline{\text{MS}}}
\def\MSBAR{\MSbar}

\def\mubar{\bar{\mu}}

\def\mubaruv{\bar{\mu}_{\mathrm{\tau_F}}}
\def\mubarir{\bar{\mu}_{_T}}

\def\mudr{\mu_{\mathrm{DR}}}
\def\muflow{\mu_\mathrm{F}}

\def\gsqms{g^2_{_{\overline{\mathrm{MS}}}}}
\def\GB{G_B}
\def\GBphys{G_B^{\text{phys.}}}

\def\gammaE{\gamma_{_{\mathrm{E}}}}

\begin{document}
\title{Quark Mass Dependence of Heavy Quark Diffusion Coefficient from Lattice QCD}
\author{Luis Altenkort}
\affiliation{Fakult\"at f\"ur Physik, Universit\"at Bielefeld, D-33615 Bielefeld, Germany}
\author{David de la Cruz}
\affiliation{Insitut f\"ur Kernphysik, Technische Universit\"at Darmstadt, Schlossgartenstra{\ss}e 2, D-64289 Darmstadt, Germany}
\author{Olaf Kaczmarek}
\affiliation{Fakult\"at f\"ur Physik, Universit\"at Bielefeld, D-33615 Bielefeld, Germany}
\author{Rasmus Larsen}
\affiliation{Department of Mathematics and Physics, University of Stavanger,
Stavanger, Norway}
\author{Guy D. Moore}
\affiliation{Insitut f\"ur Kernphysik, Technische Universit\"at Darmstadt, Schlossgartenstra{\ss}e 2, D-64289 Darmstadt, Germany}
\author{Swagato Mukherjee}
\affiliation{Physics Department, Brookhaven National Laboratory, Upton, New York 11973, USA}
\author{Peter Petreczky}
\affiliation{Physics Department, Brookhaven National Laboratory, Upton, New York 11973, USA}
\author{Hai-Tao Shu}
\email{hshu@bnl.gov}
\affiliation{Physics Department, Brookhaven National Laboratory, Upton, New York 11973, USA}
\author{Simon Stendebach}
\affiliation{Insitut f\"ur Kernphysik, Technische Universit\"at Darmstadt, Schlossgartenstra{\ss}e 2, D-64289 Darmstadt, Germany}
\collaboration{HotQCD Collaboration}
\begin{abstract}
We present the first study of the quark mass dependence of the heavy quark momentum and spatial diffusion coefficients using lattice QCD with light dynamical quarks corresponding to a pion mass of 320 MeV. We find that, for the temperature range 195 MeV $<T<$ 352 MeV, the spatial diffusion coefficients of the charm and bottom quarks are smaller than those obtained in phenomenological models that describe the $p_T$ spectra and elliptic flow of open heavy flavor hadrons. %and also has a smaller dependence on the heavy quark mass.
\end{abstract}
%
%\date{\today}
%
\maketitle
\emph{Introduction.--} 
Heavy-ion experiments at high energies hint toward a rapid thermalization of heavy (charm and bottom) quarks. These observations are surprising since the relaxation time of a heavy quark immersed in a quark-gluon plasma (QGP) is expected to be $M/T$ times larger than the relaxation time of the light bulk degrees of freedom constituting the QGP, where $M$ is the heavy quark mass~\cite{Svetitsky:1987gq,Moore:2004tg} and $T$ is the temperature of the QGP. These experimental observations corroborate the picture that the QGP created in such high-energy heavy-ion collisions is an almost perfect fluid, see Refs.~\cite{Rapp:2018qla,Dong:2019byy,He:2022ywp}. This makes the heavy quark diffusion coefficient one of the fundamental transport properties of the QGP, along with other transport coefficients such as shear and bulk viscosities. 

The heavy quark momentum diffusion coefficient $\kappa$ is defined as the average momentum transfer squared  to a heavy quark from the medium per unit time, and thus characterizes the kinetic relaxation of heavy quarks toward thermal equilibrium.
One can also define the spatial heavy quark diffusion coefficient $D_s$ in terms of the conserved net heavy flavor number through the usual Kubo formula.
This spatial heavy quark diffusion coefficient will depend on $M$, and also has a well defined limit when $M$ goes to zero. Close to equilibrium, in the limit $M \gg T$, the momentum and spatial heavy quark diffusion coefficients are related \cite{Moore:2004tg,Bouttefeux:2020ycy},
\begin{equation}
    D_s= \frac{2 T^2}{\kappa}\frac{\langle p^2 \rangle}{3MT},
    \label{eq:Ds}
\end{equation}
where $\langle p^2 \rangle$ is thermal averaged momentum squared
of the heavy quark.

Experimentally measured $p_T$ spectra and the elliptic flows of open charm and open bottom hadrons provide information on the degree of thermalization of the heavy quarks in the QGP, see Refs.~\cite{Rapp:2018qla,Dong:2019byy,He:2022ywp} for reviews.
$D_s$ can be estimated by fitting these experimental measurements using phenomenological transport models, see Refs.~\cite{Rapp:2018qla,Dong:2019byy,He:2022ywp}.
A key ingredient of these transport models is an effective momentum-dependent heavy-quark diffusion coefficient, which in turn may depend on some effective in-medium cross sections.
The $D_s$ appearing in the Kubo formula can be obtained as the zero-momentum limit of this effective diffusion coefficient.

Very recently, $D_s$ has been calculated in 2+1 flavor QCD for infinitely heavy quarks~\cite{Altenkort:2023oms}. 
This QCD result for the infinite-mass limit turns out to be smaller than
the phenomenological estimate of $D_s$ for the charm and bottom quarks.
This begs the question of whether or not these discrepancies arise solely due to the mass dependence of $D_s$.
In this Letter we address this question by presenting first lattice QCD calculations of the mass-dependent $D_s$ with light dynamical quarks.

\emph{Theoretical framework.--}
For $M\gg T$, $\kappa$ can be calculated using heavy quark effective theory~\cite{CasalderreySolana:2006rq,Caron-Huot:2009ncn,Bouttefeux:2020ycy}.
In this framework $\kappa$ is expressed in terms of the correlation function of chromo-electric ($E$) and chromo-magnetic ($B$) fields connected by fundamental Wilson lines~\cite{CasalderreySolana:2006rq,Caron-Huot:2009ncn,Bouttefeux:2020ycy}.

In this effective theory
\begin{equation}
\kappa=\kappa_E+\frac{2}{3} \langle v^2\rangle \kappa_B, 
\label{eq:kappa}
\end{equation}
where $\kappa_{E,B}(T)=2 T \lim_{\omega \to 0} \left[ \rho_{E,B}(\omega,T)/\omega \right]$~\cite{Caron-Huot:2009ncn,Bouttefeux:2020ycy}, $\rho_{E,B}$ are the spectral functions corresponding to the $E$ and $B$ field correlation functions, and $\langle v^2\rangle$ is the mean-squared thermal velocity of the heavy quark~\cite{Bouttefeux:2020ycy}. The quark mass dependence of $\kappa$ enters through $\langle v^2 \rangle$. At the leading order in $1/M$, $\langle v^2 \rangle = 3T/M$. 
In this way, $\kappa_B$ controls the quark mass dependence of $\kappa$.
%The magnitude of the quark mass dependence $\kappa$ is determined by $\kappa_B$.
% It's $3 T/M$ because of the 3 space directions.

Lattice QCD calculations of $\kappa_B$ rely on accessing $\rho_B(\omega,T)$ from the correlator~\cite{Bouttefeux:2020ycy}
\begin{equation}
    \label{eq:gblat}
    G_B(\tau, T) = \sum_{i=1}^{3} 
    \frac{\left\langle {\rm Re Tr}\left[U(\beta,\tau)B_i(\mathbf{x},\tau)U(\tau,0)B_i(\mathbf{x},0)\right]\right\rangle}{3\left\langle {\rm Re Tr} U(\beta,0)\right\rangle} \,,
\end{equation}
where $\beta=1/T$ is the inverse temperature, $\tau$ is the Euclidean time separation of the $B$ operators, and
$U(\tau_1,\tau_2)$ is a thermal Wilson line connecting the $B$-fields located at Euclidean times $\tau_1$ and $\tau_2$. 
%The $B$-fields are discretized on the lattice as
%\begin{equation}
%    \label{eq:b-field}
%B_i(\mathbf{x})\equiv\frac{\epsilon_{ijk}}{2}\left( %U_j(\mathbf{x})U_k(\mathbf{x}+\hat{j})-U_k(\mathbf{x})U_j(\mathbf{x}+\hat{k})\right).
%\end{equation}
$\rho_B$ is related to $G_B(\tau,T)$ via the integral equation
\begin{equation}
\GB(\tau,T)=\int_0^{\infty}\frac{\mathrm{d} \omega}{\pi}\ \rho_B(\omega,T) \frac{\cosh[\omega\tau-\omega/(2T))]}{\sinh[\omega/(2T)]}.
\label{eq:spectral-corr}
\end{equation}
\emph{Lattice QCD setup.--}
In the present calculation we use the same lattice QCD setup and ensembles that were used for the calculation of $\kappa_E$~\cite{Altenkort:2023oms}, specifically, 2+1 flavors of quarks in the Highly Improved Staggered Quark fermionic action~\cite{Follana:2006rc} and the tree-level improved Lüscher-Weisz gauge action~\cite{Luscher:1984xn,Luscher:1985zq} with physical values of the kaon and 320~MeV pion masses and at $T=$195, 220, 251, 293 and 352 MeV. At each temperature (except 352 MeV) we use three lattice spacings ($a$) to carry out continuum extrapolations ($a\to0$) of $\GB$. Further details are provided in Supplemental Material~\cite{supplemental}.

The $B$-fields are discretized on the lattice as $B_i(\mathbf{x})\equiv \epsilon_{ijk}\left( U_j(\mathbf{x})U_k(\mathbf{x}+\hat{j})-U_k(\mathbf{x})U_j(\mathbf{x}+\hat{k})\right)/2.$ For measurements of $G_B$ we use a Symanzik-improved version~\cite{Ramos:2015baa} of gradient flow~\cite{Narayanan:2006rf,Luscher:2009eq,Luscher:2010iy,Luscher:2011bx}. Guided by our experience~\cite{Altenkort:2020fgs,Brambilla:2022xbd} and perturbative QCD (pQCD)~\cite{Eller:2018yje} we limit the gradient-flow time ($\tauf$) within the range $\sqrt{8 \tauf} < \tau/3$. We find that gradient flow improves the signal-to-noise ratio of $\GB$.

At 1-loop level~\cite{Eichten:1990vp} in pQCD $\GB$ has a nontrivial anomalous dimension, and gradient flow serves as a nonperturbative renormalization scheme for $\GB$.
The continuum-extrapolated $\GB$ are renormalized in the gradient-flow scheme at the scale $\muflow=1/\sqrt{8\tauf}$. The renormalization-group invariant physical correlator,  $\GBphys$, is obtained via the one-loop pQCD matching~\cite{guy-david-matching}
\begin{equation}
    \GBphys(\tau,T) = \lim_{\tauf \rightarrow 0} 
    \Zmatch(\mubarir,\mubaruv,\muflow) \, \GB(\tau,T,\tauf).
\label{eq:match-to-physical}
\end{equation}
The power-law corrections arising from mixing with high-dimension operators are removed through the $\tauf\to 0$ extrapolations at each $T$. The matching function, $\Zmatch$, involves three components: matching from the gradient-flow to the $\MSBAR$ scheme at a scale $\mubaruv$,
matching between $\MSBAR$-renormalized thermal QCD to the static quark effective theory at a scale $\mubarir$, and running of the anomalous dimension of the operator from  $\mubarir$ to  $\mubaruv$. If $\Zmatch$ were known up to all orders in pQCD, the dependence of $\GBphys$ on the scales $\mubarir$ and $\mubaruv$ would exactly cancel.
Since $\Zmatch$ is known only up to one loop (NLO), we estimate the uncertainty from unknown higher-order effects in the matching by varying the values of each scale; for $\mubarir$ we consider the two choices $\mubarir=2\pi T$ or 19.18$T$, and for $\mubaruv$ we consider $\mubaruv=\muflow$ or 1.4986$\muflow$. Further details and the expression for $\Zmatch$ are given in Supplemental Material~\cite{supplemental}.

\begin{figure*}[!th]
\centering
\includegraphics[width=0.45\textwidth,height=0.33\textheight]{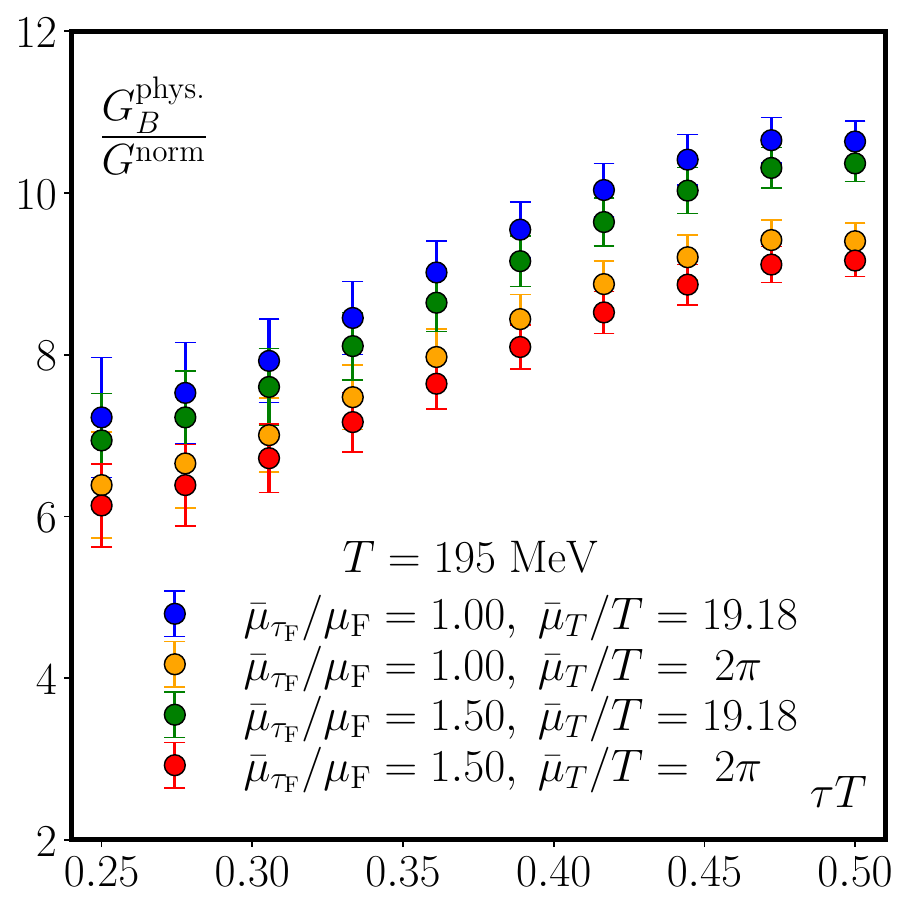}
\includegraphics[width=0.45\textwidth,height=0.33\textheight]{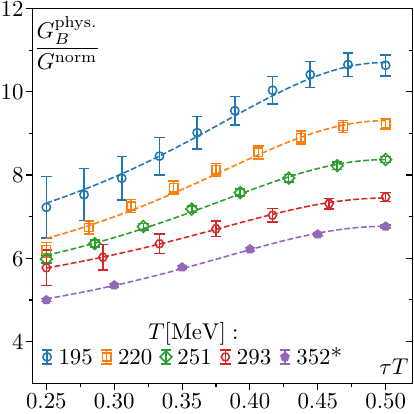}
\caption{Left: Scale dependence of $\GBphys$ [cf Eq.~(\ref{eq:match-to-physical})] at $T=195$ MeV. Right:
Temperature dependence of $\GBphys$ for $\mubarir/T=19.18$ and $\mubaruv/\muflow=1.0$. The dashed curves denote the central values of the $smax$-model fit to $\GBphys$ with NLO $\rho_{\mathrm{uv}}$ at scale $\mu=\sqrt{(0.13306\omega)^2+\mudr^2}$.}
\label{fig:corrs-fit}
\end{figure*}
\begin{figure}[!h]
\centering
\vspace*{-0.5cm}
\includegraphics[width=0.48\textwidth,height=0.33\textheight]{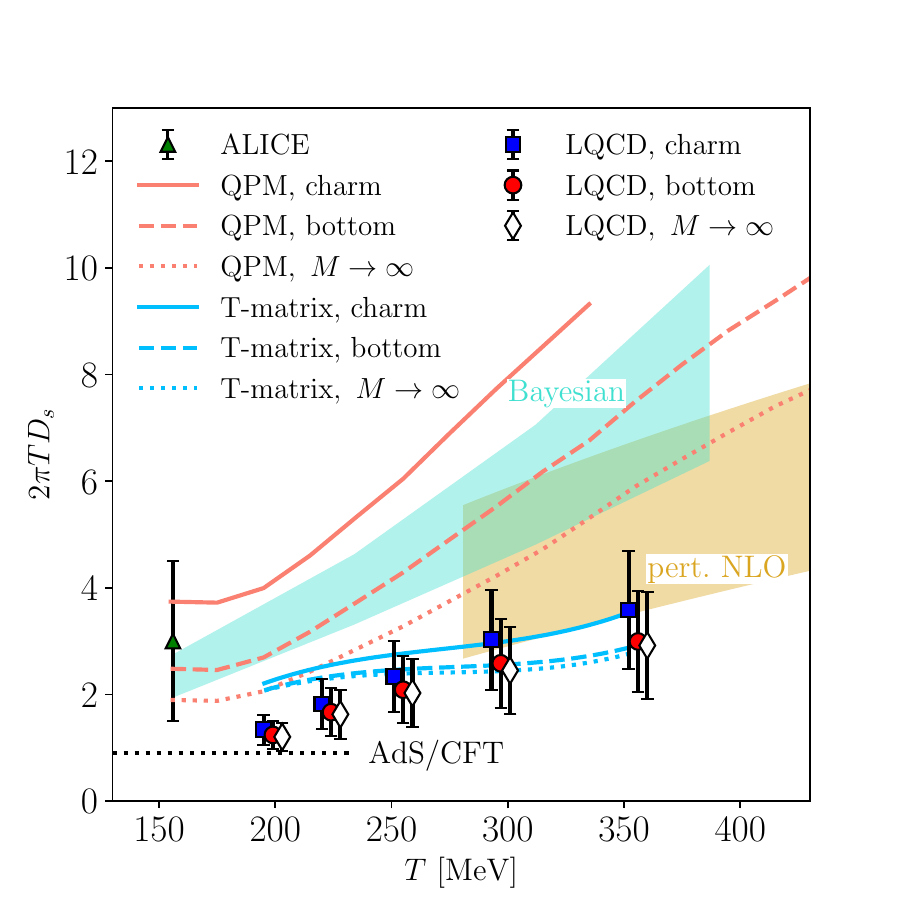}
\vspace*{-0.5cm}
\caption{Lattice QCD (LQCD) results for the spatial diffusion coefficients, $D_s$, for the charm, bottom and infinitely-heavy quarks compared with those from the quasi-particle model (QPM)~\cite{Sambataro:2023tlv} and the T-matrix approach~\cite{Liu:2016ysz, ZhanduoTang:2023ewm}. Also shown are the infinitely-heavy quark diffusion coefficients from the ALICE collaboration's phenomenological estimate~\cite{Xu:2017obm,ALICE:2021rxa}, NLO perturbative calculation~\cite{CaronHuot:2007gq} and AdS/CFT estimate at a certain value of $\lambda$~\cite{CasalderreySolana:2006rq}. The width of the NLO perturbative QCD band corresponds to the variation of the renormalization scale from $\mu=2 \pi T$ (upper boundary of the orange band) and $\mu=4 \pi T$ (lower boundary).}
\label{fig:summary}
\vspace*{-0.5cm}
\end{figure}
\emph{Data analysis.--}

The $B$-field correlator $G_B$ scales with a strong negative power of $\tau$ since $\rho_B\propto\omega^3$ at large $\omega$. To mitigate this as well as the lattice artifacts and distortion due to gradient
flow~\cite{Altenkort:2023oms} we normalize $G_B$ with $G^{\mathrm{norm}}(\tau,T,N_\tau,\tauf)=G_B^{\mathrm{LO}}(\tau,T,N_\tau,\tauf)/(C_F g^2)$, where $G_B^{\mathrm{LO}}$ is the tree-level pQCD results for $G_B$ at nonzero lattice spacing and gradient-flow time, $C_F$ is the Casimir factor, and $g^2$ is the strong coupling. For brevity we suppress the arguments of the normalization correlator.

Our lattice data are not $G_B(\tau,T,\tauf)$ directly, but
$G_B(\tau,T,N_\tau,\tauf)$ with $N_{\tau}$ the number of points across the lattice time direction, which is related to the lattice spacing via $N_\tau a = 1/T$.
Therefore, before performing the $\tauf\to 0$ limit in Eq.~(\ref{eq:match-to-physical}), we must first take the $N_{\tau} \to \infty$ limit to find $\GB(\tau,T,\tauf)$.
Because the available $\tau$ values also depend on the lattice spacing, we first perform a $\tau$-interpolation of $\GB(\tau,T,N_{\tau},\tauf)/G^{\mathrm{norm}}$ on the two coarsest lattices, separately at each $(T,\tauf)$ pair, to establish the value at the $\tau$ values available on the finest lattice.
This then allows the $N_{\tau} \to \infty$ extrapolation,
which is performed independently at each $(\tau,T,\tauf)$ triple.
In extrapolating to $N_{\tau} \to \infty$ we assume that the discretization errors scale as $1/N_{\tau}^2 =(aT)^2$. 

For each $\tau T$, to take the $\tauf\rightarrow 0$ limit we multiply the continuum-extrapolated results for $\GB/G^{\mathrm{norm}}$ by $\Zmatch$ and do linear extrapolations in $\tauf$, as suggested by NLO pQCD~\cite{Eller:2021qpp}. To avoid potentially large  discretization effects and to keep the gradient-flow scale smaller than all relevant physical scales in the problem, we restrict to the range of flow times $0.25\leq \sqrt{8\tauf}/\tau\leq 0.30$~\cite{Altenkort:2023oms}. At $T=352$ MeV we also perform such flow extrapolation based on the single lattice.

As an example, in Fig.~\ref{fig:corrs-fit} (left panel) we show the resulting $\GBphys$ for different choices of $\mubaruv$ and $\mubarir$. In Fig.~\ref{fig:corrs-fit} (right panel) we show the $T$ dependence of $\GBphys$ for $\mubarir/T=19.18$ and $\mubaruv/\muflow=1.0$. Further details on and examples of $\GBphys$ can be found in Supplemental Material~\cite{supplemental}.

%The double-extrapolated correlators are shown in the left panel of Fig.~\ref{fig:corrs-fit} for $T=195$ MeV as an example to illustrate the different choices of $\mubaruv$ and $\mubarir$. In the right panel of Fig.~\ref{fig:corrs-fit} we show the double-extrapolated normalized correlators for all temperatures obtained with $\mubarir/T=19.18$ and $\mubaruv/\muflow=1.0$. More details about the double extrapolation can be found in the Supplemental Material~\cite{supplemental}. As one can see from the figure, there is a slight dependence of $\GBphys/\Gnorm$ on the the choice of $\mubarir$ and $\mubaruv$. The $\tau$ and the temperature dependence ofthe $B$-field correlator is very similar to that of the chromo-electric correlator \cite{Altenkort:2023oms}.

$\GBphys$ are fitted to Eq.~(\ref{eq:spectral-corr}) to obtain $\kappa_B$. $\kappa_B$ is encoded in the infrared region of the spectral function, $\rho^{\mathrm{ir}}_B(\omega) = \omega \kappa_B/(2T)$. 
For the ultraviolet region of spectral function, $\rho^{\mathrm{uv}}_B(\omega,\mu)$, we use leading order and NLO vacuum pQCD results. To convert the NLO pQCD results~\cite{Banerjee:2022uge,guy-david-matching} from the $\MSBAR$ scheme, at the scale $\mu$, to the renormalization-group invariant physical scheme we use the matching NLO Wilson coefficient~\cite{Laine:2021uzs}, $c_B^2(\mu,\mubarir)$, to the static quark effective theory. To account for possible higher order effects we also introduce another (in addition to $\kappa_B$) $\omega$-independent fit parameter, $K$. 
The resulting choices for the UV part of the physical spectral function are $\rho^{\mathrm{uv,phys.}}_B(\omega)=\left\{K\rho^{\mathrm{uv,LO}}_B(\omega,\mu),\ K c_B^2(\mu,\mubarir)\rho^{\mathrm{uv,NLO}}_B(\omega,\mu)\right\}$. 
% \textbf{\color{red}Luis: I completed a sentence here which was incomplete before}
Expressions for $\rho^{\mathrm{uv,LO}}$, $\rho^{\mathrm{uv,NLO}}$, and  $c_B^2(\mu,\mubarir)$ are given in the Supplemental Material~\cite{supplemental}.

Following the previous works~\cite{Francis:2015daa,Altenkort:2020fgs,Brambilla:2020siz,Altenkort:2023oms} we use three models to interpolate between $\rho^{\mathrm{ir}}_B$ and $\rho^{\mathrm{uv}}_B$ to obtain $\rho_B$ over all $\omega$. (1) The maximum ($max$) model, $\rho_{\mathrm{max}}=\max \Big{(}\rho_B^\mathrm{ir}(\omega),\ \rho_B^\mathrm{uv,phys.}(\omega)\Big{)}$, which imposes a hard switch-over between the two regimes. (2) The smooth maximum ($smax$) model, $\rho_{\mathrm{smax}}=\sqrt{(\rho_B^\mathrm{ir}(\omega))^2+(\rho_B^\mathrm{uv,phys.}(\omega))^2}$, which imposes a smooth switch-over. (3) The power-law ($plaw$) model, $\rho_{\mathrm{plaw}}$, which is $\rho_B^\mathrm{ir}(\omega)$ up to $\omega=\omega_{\mathrm{ir}}$ and $\rho_B^\mathrm{uv,phys}(\omega)$ above $\omega=\omega^\mathrm{uv}$. In between, it is connected by a power-law curve $\rho_B(\omega)=c\ \omega^p$. The parameters $c$ and $p$ are chosen to provide continuity at the boundary.
Physically motivated choices of these boundaries are $\omega_\mathrm{ir}=T$ and $\omega_\mathrm{uv}=2 \pi T$~\cite{Altenkort:2023oms}. 

$\rho_{\mathrm{max}}$, $\rho_{\mathrm{smax}}$, and $\rho_{\mathrm{plaw}}$ also depend on the intermediate $\MSBAR$ renormalization scale $\mu$. For reasons discussed in the Supplemental Material~\cite{supplemental}, we consider two options $\mu=\sqrt{(0.13306\omega)^2+\mudr^2}$ and $\mu=\sqrt{4\omega^2+\mudr^2}$, where $\mudr \approx 9.1T$ is a typical thermal scale inferred from the high temperature three dimensional effective theory~\cite{Kajantie:1997tt}.

\emph{Results.---}
Combining different choices of (1) $\mubarir$ and $\mubaruv$ in $\GBphys$, (2) $\rho^{\mathrm{uv,phys.}}_B(\omega, \mu)$ and $\mu$, and (3) three interpolating models $\rho_{\mathrm{max}}$, $\rho_{\mathrm{smax}}$, and $\rho_{\mathrm{plaw}}$, we carry out 24 different fits for $K$ and $\kappa_B$ on each bootstrap sample of gauge configurations at each $T$. The final result for $\kappa_B$ at each $T$ is obtained from the median and 68\% confidence limit of the distribution of all the bootstrap samples over gauge configurations and fit forms and, thus, include both statistical as well as systematic errors arising from different model and scales choices. We find $\kappa_B(T=195\,\mathrm{MeV})= 10.42_{-3.61}^{+2.66} T^3$, $\kappa_B(T=220\,\mathrm{MeV})= 8.78_{-3.21}^{+2.20} T^3$, $\kappa_B(T=251\,\mathrm{MeV})= 7.18_{-2.90}^{+1.94}T^3$, $\kappa_B(T=293\,\mathrm{MeV})= 5.02_{-2.24}^{+1.92} T^3$ and $\kappa_B(T=352\,\mathrm{MeV})= 5.66_{-2.41}^{+1.50} T^3$, which are of similar magnitude to $\kappa_E$ obtained for the same ensembles~\cite{Altenkort:2023oms}. The $\kappa_B$ for 2+1 flavor QCD turns out be much larger than those from quenched QCD~\cite{Banerjee:2022uge,Brambilla:2022xbd} at the same values of $T/T_c$. For further details on the fits and results see Supplemental Material~\cite{supplemental}.

The $\langle v^2\rangle$ appearing in Eq.~(\ref{eq:kappa}) can be obtained either from the low-frequency part of the spectral function corresponding to the net-flavor current~\cite{Caron-Huot:2009ncn,Burnier:2012ze} or in
the quasi-particle model with a temperature
dependent quark mass ~\cite{Petreczky:2008px}. In quenched QCD it
has been shown that a quasi-particle model with a temperature dependent
heavy quark mass fitted to the heavy quark number susceptibility 
gives a $\langle v^2\rangle$ that agrees with the one obtained
from the low-frequency part of the net heavy quark current
spectral function ~\cite{Petreczky:2008px}. Therefore, in this
work we adopt the quasi-particle model to calculate 
$\langle v^2 \rangle$.
For the temperature dependent charm quark is obtained from the  continuum-extrapolated lattice QCD results for the charm susceptibility~\cite{Bellwied:2015lba}. No lattice QCD results for bottom quark susceptibility are available presently. Therefore, we simply fix the effective bottom quark mass to $4.8$ GeV. The $\langle p^2 \rangle$ needed to obtain $D_s$, cf. Eq.~(\ref{eq:Ds}), are estimated from the quasi-particle model in the same way as for the $\langle v^2 \rangle$. For further details, see Supplemental Material~\cite{supplemental}. 
There we also show that the $\langle v^2 \rangle$, $\langle p^2 \rangle$
and the values of $D_s$ are not too sensitive to the precise choice
of the bottom quark mass.

Final results for $D_s$ are summarized in Fig.~\ref{fig:summary}. We find a slight increase in $D_s$ with decreasing heavy quark mass. We compare our results with those obtained from a phenomenological quasi-particle model (QMP)~\cite{Sambataro:2023tlv} and the T-matrix approach~\cite{Liu:2016ysz, ZhanduoTang:2023ewm}. We find that $D_s$ in QCD is smaller than the results of these calculations and shows smaller dependence on the heavy quark mass, with the exception of the T-matrix result at
the highest temperature considered by us. 
Our result for $D_s$ is also smaller than other phenomenological estimates \cite{Xu:2017obm,ALICE:2021rxa}, which do not take into account the quark mass dependence. Finally, for completeness, we show the AdS/CFT estimate~\cite{CasalderreySolana:2006rq} of $D_s$ with a certain value of $\lambda$ and the result of the NLO perturbative calculation~\cite{CaronHuot:2007gq} in the limit $M \rightarrow \infty$.

\emph{Conclusion.--}
We presented the 2+1 flavor lattice QCD calculations of the quark mass dependence of the momentum and spatial heavy quark diffusion coefficients. In the temperature range 195~MeV~$\le T \le$~352~MeV the quark mass dependence turns out to be quite small. In conjunction with the previous results for an infinitely-heavy quark~\cite{Altenkort:2023oms}, the present calculations provide the first non-perturbative QCD results for the charm and bottom quark diffusion coefficients in the QGP. These non-perturbative QCD results will serve as critical inputs to and benchmarks for various dynamical models to study thermalization of charm and bottom quarks in the strongly-coupled medium created in heavy-ion collisions at the Large Hadron Collider and the Relativistic Heavy-Ion Collider.

All computations in this work were performed using \texttt{SIMULATeQCD}~\cite{Mazur:2021zgi, Bollweg:2021cvl, HotQCD:2023ghu}.

All data from our calculations, presented in the figures of this paper, can be found in \cite{datapublication}.

\emph{Acknowledgments.---}
This material is based upon work supported by the U.S. Department of Energy, Office of Science, Office of Nuclear Physics through Contract No.~DE-SC0012704, and within the frameworks of Scientific Discovery through Advanced Computing (SciDAC) award \textit{Fundamental Nuclear Physics at the Exascale and Beyond} and the Topical Collaboration in Nuclear Theory \textit{Heavy-Flavor Theory (HEFTY) for QCD Matter}. This work is supported by the Deutsche For\-schungs\-ge\-mein\-schaft
(DFG, German Research Foundation) through the CRC-TR 211 ``Strong-interaction matter under extreme conditions''– Project No. 315477589 – TRR 211. R.~L. acknowledges funding by the Research Council of Norway under the FRIPRO Young Research Talent Grant No. 286883. 
We thank Szabolcs Bors\'anyi for providing the continuum extrapolated
data for charm susceptibility.

This research used awards of computer time provided by: the ALCC program at the Oak Ridge Leadership Computing Facility, which is a DOE Office of Science User Facility supported under Contract No. DE-AC05-00OR22725; the National Energy Research Scientific Computing Center (NERSC), a U.S. Department of Energy Office of Science User Facility located at Lawrence Berkeley National Laboratory, operated under Contract No. DE-AC02- 05CH11231; 
%\SM{Euro-HPC time on LUMI-G}; 
the PRACE awards on JUWELS at GCS@FZJ, Germany and Marconi100 at CINECA, Italy. Computations for this work were carried out in part on facilities of the USQCD Collaboration, which are funded by the Office of Science of the U.S. Department of Energy. Parts of the computations in this work also were performed at Bielefeld University's
GPU Cluster, supported by HPC.NRW.

\bibliographystyle{apsrev4-1}
\bibliography{references}

%merlin.mbs apsrev4-1.bst 2010-07-25 4.21a (PWD, AO, DPC) hacked
%Control: key (0)
%Control: author (72) initials jnrlst
%Control: editor formatted (1) identically to author
%Control: production of article title (-1) disabled
%Control: page (0) single
%Control: year (1) truncated
%Control: production of eprint (0) enabled
\begin{thebibliography}{54}%
\makeatletter
\providecommand \@ifxundefined [1]{%
 \@ifx{#1\undefined}
}%
\providecommand \@ifnum [1]{%
 \ifnum #1\expandafter \@firstoftwo
 \else \expandafter \@secondoftwo
 \fi
}%
\providecommand \@ifx [1]{%
 \ifx #1\expandafter \@firstoftwo
 \else \expandafter \@secondoftwo
 \fi
}%
\providecommand \natexlab [1]{#1}%
\providecommand \enquote  [1]{``#1''}%
\providecommand \bibnamefont  [1]{#1}%
\providecommand \bibfnamefont [1]{#1}%
\providecommand \citenamefont [1]{#1}%
\providecommand \href@noop [0]{\@secondoftwo}%
\providecommand \href [0]{\begingroup \@sanitize@url \@href}%
\providecommand \@href[1]{\@@startlink{#1}\@@href}%
\providecommand \@@href[1]{\endgroup#1\@@endlink}%
\providecommand \@sanitize@url [0]{\catcode `\\12\catcode `\$12\catcode
  `\&12\catcode `\#12\catcode `\^12\catcode `\_12\catcode `\%12\relax}%
\providecommand \@@startlink[1]{}%
\providecommand \@@endlink[0]{}%
\providecommand \url  [0]{\begingroup\@sanitize@url \@url }%
\providecommand \@url [1]{\endgroup\@href {#1}{\urlprefix }}%
\providecommand \urlprefix  [0]{URL }%
\providecommand \Eprint [0]{\href }%
\providecommand \doibase [0]{http://dx.doi.org/}%
\providecommand \selectlanguage [0]{\@gobble}%
\providecommand \bibinfo  [0]{\@secondoftwo}%
\providecommand \bibfield  [0]{\@secondoftwo}%
\providecommand \translation [1]{[#1]}%
\providecommand \BibitemOpen [0]{}%
\providecommand \bibitemStop [0]{}%
\providecommand \bibitemNoStop [0]{.\EOS\space}%
\providecommand \EOS [0]{\spacefactor3000\relax}%
\providecommand \BibitemShut  [1]{\csname bibitem#1\endcsname}%
\let\auto@bib@innerbib\@empty
%</preamble>
\bibitem [{\citenamefont {Svetitsky}(1988)}]{Svetitsky:1987gq}%
  \BibitemOpen
  \bibfield  {author} {\bibinfo {author} {\bibfnamefont {B.}~\bibnamefont
  {Svetitsky}},\ }\href {\doibase 10.1103/PhysRevD.37.2484} {\bibfield
  {journal} {\bibinfo  {journal} {Phys. Rev. D}\ }\textbf {\bibinfo {volume}
  {37}},\ \bibinfo {pages} {2484} (\bibinfo {year} {1988})}\BibitemShut
  {NoStop}%
\bibitem [{\citenamefont {Moore}\ and\ \citenamefont
  {Teaney}(2005)}]{Moore:2004tg}%
  \BibitemOpen
  \bibfield  {author} {\bibinfo {author} {\bibfnamefont {G.~D.}\ \bibnamefont
  {Moore}}\ and\ \bibinfo {author} {\bibfnamefont {D.}~\bibnamefont {Teaney}},\
  }\href {\doibase 10.1103/PhysRevC.71.064904} {\bibfield  {journal} {\bibinfo
  {journal} {Phys. Rev. C}\ }\textbf {\bibinfo {volume} {71}},\ \bibinfo
  {pages} {064904} (\bibinfo {year} {2005})},\ \Eprint
  {http://arxiv.org/abs/hep-ph/0412346} {arXiv:hep-ph/0412346} \BibitemShut
  {NoStop}%
\bibitem [{\citenamefont {Beraudo}\ \emph {et~al.}(2018)\citenamefont {Beraudo}
  \emph {et~al.}}]{Rapp:2018qla}%
  \BibitemOpen
  \bibfield  {author} {\bibinfo {author} {\bibfnamefont {A.}~\bibnamefont
  {Beraudo}} \emph {et~al.},\ }\href {\doibase 10.1016/j.nuclphysa.2018.09.002}
  {\bibfield  {journal} {\bibinfo  {journal} {Nucl. Phys. A}\ }\textbf
  {\bibinfo {volume} {979}},\ \bibinfo {pages} {21} (\bibinfo {year} {2018})},\
  \Eprint {http://arxiv.org/abs/1803.03824} {arXiv:1803.03824 [nucl-th]}
  \BibitemShut {NoStop}%
\bibitem [{\citenamefont {Dong}\ \emph {et~al.}(2019)\citenamefont {Dong},
  \citenamefont {Lee},\ and\ \citenamefont {Rapp}}]{Dong:2019byy}%
  \BibitemOpen
  \bibfield  {author} {\bibinfo {author} {\bibfnamefont {X.}~\bibnamefont
  {Dong}}, \bibinfo {author} {\bibfnamefont {Y.-J.}\ \bibnamefont {Lee}}, \
  and\ \bibinfo {author} {\bibfnamefont {R.}~\bibnamefont {Rapp}},\ }\href
  {\doibase 10.1146/annurev-nucl-101918-023806} {\bibfield  {journal} {\bibinfo
   {journal} {Ann. Rev. Nucl. Part. Sci.}\ }\textbf {\bibinfo {volume} {69}},\
  \bibinfo {pages} {417} (\bibinfo {year} {2019})},\ \Eprint
  {http://arxiv.org/abs/1903.07709} {arXiv:1903.07709 [nucl-ex]} \BibitemShut
  {NoStop}%
\bibitem [{\citenamefont {He}\ \emph {et~al.}(2023)\citenamefont {He},
  \citenamefont {van Hees},\ and\ \citenamefont {Rapp}}]{He:2022ywp}%
  \BibitemOpen
  \bibfield  {author} {\bibinfo {author} {\bibfnamefont {M.}~\bibnamefont
  {He}}, \bibinfo {author} {\bibfnamefont {H.}~\bibnamefont {van Hees}}, \ and\
  \bibinfo {author} {\bibfnamefont {R.}~\bibnamefont {Rapp}},\ }\href {\doibase
  10.1016/j.ppnp.2023.104020} {\bibfield  {journal} {\bibinfo  {journal} {Prog.
  Part. Nucl. Phys.}\ }\textbf {\bibinfo {volume} {130}},\ \bibinfo {pages}
  {104020} (\bibinfo {year} {2023})},\ \Eprint
  {http://arxiv.org/abs/2204.09299} {arXiv:2204.09299 [hep-ph]} \BibitemShut
  {NoStop}%
\bibitem [{\citenamefont {Bouttefeux}\ and\ \citenamefont
  {Laine}(2020)}]{Bouttefeux:2020ycy}%
  \BibitemOpen
  \bibfield  {author} {\bibinfo {author} {\bibfnamefont {A.}~\bibnamefont
  {Bouttefeux}}\ and\ \bibinfo {author} {\bibfnamefont {M.}~\bibnamefont
  {Laine}},\ }\href {\doibase 10.1007/JHEP12(2020)150} {\bibfield  {journal}
  {\bibinfo  {journal} {JHEP}\ }\textbf {\bibinfo {volume} {12}},\ \bibinfo
  {pages} {150} (\bibinfo {year} {2020})},\ \Eprint
  {http://arxiv.org/abs/2010.07316} {arXiv:2010.07316 [hep-ph]} \BibitemShut
  {NoStop}%
\bibitem [{\citenamefont {Altenkort}\ \emph
  {et~al.}(2023{\natexlab{a}})\citenamefont {Altenkort}, \citenamefont
  {Kaczmarek}, \citenamefont {Larsen}, \citenamefont {Mukherjee}, \citenamefont
  {Petreczky}, \citenamefont {Shu},\ and\ \citenamefont
  {Stendebach}}]{Altenkort:2023oms}%
  \BibitemOpen
  \bibfield  {author} {\bibinfo {author} {\bibfnamefont {L.}~\bibnamefont
  {Altenkort}}, \bibinfo {author} {\bibfnamefont {O.}~\bibnamefont
  {Kaczmarek}}, \bibinfo {author} {\bibfnamefont {R.}~\bibnamefont {Larsen}},
  \bibinfo {author} {\bibfnamefont {S.}~\bibnamefont {Mukherjee}}, \bibinfo
  {author} {\bibfnamefont {P.}~\bibnamefont {Petreczky}}, \bibinfo {author}
  {\bibfnamefont {H.-T.}\ \bibnamefont {Shu}}, \ and\ \bibinfo {author}
  {\bibfnamefont {S.}~\bibnamefont {Stendebach}} (\bibinfo {collaboration}
  {HotQCD}),\ }\href {\doibase 10.1103/PhysRevLett.130.231902} {\bibfield
  {journal} {\bibinfo  {journal} {Phys. Rev. Lett.}\ }\textbf {\bibinfo
  {volume} {130}},\ \bibinfo {pages} {231902} (\bibinfo {year}
  {2023}{\natexlab{a}})},\ \Eprint {http://arxiv.org/abs/2302.08501}
  {arXiv:2302.08501 [hep-lat]} \BibitemShut {NoStop}%
\bibitem [{\citenamefont {Casalderrey-Solana}\ and\ \citenamefont
  {Teaney}(2006)}]{CasalderreySolana:2006rq}%
  \BibitemOpen
  \bibfield  {author} {\bibinfo {author} {\bibfnamefont {J.}~\bibnamefont
  {Casalderrey-Solana}}\ and\ \bibinfo {author} {\bibfnamefont
  {D.}~\bibnamefont {Teaney}},\ }\href {\doibase 10.1103/PhysRevD.74.085012}
  {\bibfield  {journal} {\bibinfo  {journal} {Phys. Rev. D}\ }\textbf {\bibinfo
  {volume} {74}},\ \bibinfo {pages} {085012} (\bibinfo {year} {2006})},\
  \Eprint {http://arxiv.org/abs/hep-ph/0605199} {arXiv:hep-ph/0605199}
  \BibitemShut {NoStop}%
\bibitem [{\citenamefont {Caron-Huot}\ \emph {et~al.}(2009)\citenamefont
  {Caron-Huot}, \citenamefont {Laine},\ and\ \citenamefont
  {Moore}}]{Caron-Huot:2009ncn}%
  \BibitemOpen
  \bibfield  {author} {\bibinfo {author} {\bibfnamefont {S.}~\bibnamefont
  {Caron-Huot}}, \bibinfo {author} {\bibfnamefont {M.}~\bibnamefont {Laine}}, \
  and\ \bibinfo {author} {\bibfnamefont {G.~D.}\ \bibnamefont {Moore}},\ }\href
  {\doibase 10.1088/1126-6708/2009/04/053} {\bibfield  {journal} {\bibinfo
  {journal} {JHEP}\ }\textbf {\bibinfo {volume} {04}},\ \bibinfo {pages} {053}
  (\bibinfo {year} {2009})},\ \Eprint {http://arxiv.org/abs/0901.1195}
  {arXiv:0901.1195 [hep-lat]} \BibitemShut {NoStop}%
\bibitem [{\citenamefont {Follana}\ \emph {et~al.}(2007)\citenamefont
  {Follana}, \citenamefont {Mason}, \citenamefont {Davies}, \citenamefont
  {Hornbostel}, \citenamefont {Lepage}, \citenamefont {Shigemitsu},
  \citenamefont {Trottier},\ and\ \citenamefont {Wong}}]{Follana:2006rc}%
  \BibitemOpen
  \bibfield  {author} {\bibinfo {author} {\bibfnamefont {E.}~\bibnamefont
  {Follana}}, \bibinfo {author} {\bibfnamefont {Q.}~\bibnamefont {Mason}},
  \bibinfo {author} {\bibfnamefont {C.}~\bibnamefont {Davies}}, \bibinfo
  {author} {\bibfnamefont {K.}~\bibnamefont {Hornbostel}}, \bibinfo {author}
  {\bibfnamefont {G.~P.}\ \bibnamefont {Lepage}}, \bibinfo {author}
  {\bibfnamefont {J.}~\bibnamefont {Shigemitsu}}, \bibinfo {author}
  {\bibfnamefont {H.}~\bibnamefont {Trottier}}, \ and\ \bibinfo {author}
  {\bibfnamefont {K.}~\bibnamefont {Wong}} (\bibinfo {collaboration} {HPQCD,
  UKQCD}),\ }\href {\doibase 10.1103/PhysRevD.75.054502} {\bibfield  {journal}
  {\bibinfo  {journal} {Phys. Rev.}\ }\textbf {\bibinfo {volume} {D75}},\
  \bibinfo {pages} {054502} (\bibinfo {year} {2007})},\ \Eprint
  {http://arxiv.org/abs/hep-lat/0610092} {arXiv:hep-lat/0610092 [hep-lat]}
  \BibitemShut {NoStop}%
%%CITATION = HEP-LAT/0610092;%%
\bibitem [{\citenamefont {Luscher}\ and\ \citenamefont
  {Weisz}(1985{\natexlab{a}})}]{Luscher:1984xn}%
  \BibitemOpen
  \bibfield  {author} {\bibinfo {author} {\bibfnamefont {M.}~\bibnamefont
  {Luscher}}\ and\ \bibinfo {author} {\bibfnamefont {P.}~\bibnamefont
  {Weisz}},\ }\href {\doibase 10.1007/BF01206178} {\bibfield  {journal}
  {\bibinfo  {journal} {Commun. Math. Phys.}\ }\textbf {\bibinfo {volume}
  {97}},\ \bibinfo {pages} {59} (\bibinfo {year} {1985}{\natexlab{a}})},\
  \bibinfo {note} {[Erratum: Commun.Math.Phys. 98, 433 (1985)]}\BibitemShut
  {NoStop}%
\bibitem [{\citenamefont {Luscher}\ and\ \citenamefont
  {Weisz}(1985{\natexlab{b}})}]{Luscher:1985zq}%
  \BibitemOpen
  \bibfield  {author} {\bibinfo {author} {\bibfnamefont {M.}~\bibnamefont
  {Luscher}}\ and\ \bibinfo {author} {\bibfnamefont {P.}~\bibnamefont
  {Weisz}},\ }\href {\doibase 10.1016/0370-2693(85)90966-9} {\bibfield
  {journal} {\bibinfo  {journal} {Phys. Lett. B}\ }\textbf {\bibinfo {volume}
  {158}},\ \bibinfo {pages} {250} (\bibinfo {year}
  {1985}{\natexlab{b}})}\BibitemShut {NoStop}%
\bibitem [{sup()}]{supplemental}%
  \BibitemOpen
  \href@noop {} {\bibinfo  {journal} {See Supplemental Material for the
  technical details of this study, which includes Refs.~\cite{2022-Stendebach,
  Altenkort:2020fgs, Altenkort:2023oms, Banerjee:2022gen, Banerjee:2022uge,
  Bazavov:2010sb, Bazavov:2017dsy, Bellwied:2015lba, Brambilla:2020siz,
  Brambilla:2022xbd, Caron-Huot:2009ncn, Chetyrkin:2000yt, Clark:2003na,
  Clark:2004cp, FlavourLatticeAveragingGroupFLAG:2021npn, Herren:2017osy,
  HotQCD:2014kol, Kajantie:1997tt, Laine:2021uzs, MILC:2010hzw,
  Petreczky:2008px, guy-david-matching}}\ }\BibitemShut {NoStop}%
\bibitem [{\citenamefont {Ramos}\ and\ \citenamefont
  {Sint}(2016)}]{Ramos:2015baa}%
  \BibitemOpen
\bibfield  {journal} {  }\bibfield  {author} {\bibinfo {author} {\bibfnamefont
  {A.}~\bibnamefont {Ramos}}\ and\ \bibinfo {author} {\bibfnamefont
  {S.}~\bibnamefont {Sint}},\ }\href {\doibase 10.1140/epjc/s10052-015-3831-9}
  {\bibfield  {journal} {\bibinfo  {journal} {Eur. Phys. J. C}\ }\textbf
  {\bibinfo {volume} {76}},\ \bibinfo {pages} {15} (\bibinfo {year} {2016})},\
  \Eprint {http://arxiv.org/abs/1508.05552} {arXiv:1508.05552 [hep-lat]}
  \BibitemShut {NoStop}%
\bibitem [{\citenamefont {Narayanan}\ and\ \citenamefont
  {Neuberger}(2006)}]{Narayanan:2006rf}%
  \BibitemOpen
  \bibfield  {author} {\bibinfo {author} {\bibfnamefont {R.}~\bibnamefont
  {Narayanan}}\ and\ \bibinfo {author} {\bibfnamefont {H.}~\bibnamefont
  {Neuberger}},\ }\href {\doibase 10.1088/1126-6708/2006/03/064} {\bibfield
  {journal} {\bibinfo  {journal} {JHEP}\ }\textbf {\bibinfo {volume} {03}},\
  \bibinfo {pages} {064} (\bibinfo {year} {2006})},\ \Eprint
  {http://arxiv.org/abs/hep-th/0601210} {arXiv:hep-th/0601210 [hep-th]}
  \BibitemShut {NoStop}%
%%CITATION = HEP-TH/0601210;%%
\bibitem [{\citenamefont {Lüscher}(2010)}]{Luscher:2009eq}%
  \BibitemOpen
  \bibfield  {author} {\bibinfo {author} {\bibfnamefont {M.}~\bibnamefont
  {Lüscher}},\ }\href {\doibase 10.1007/s00220-009-0953-7} {\bibfield
  {journal} {\bibinfo  {journal} {Commun. Math. Phys.}\ }\textbf {\bibinfo
  {volume} {293}},\ \bibinfo {pages} {899} (\bibinfo {year} {2010})},\ \Eprint
  {http://arxiv.org/abs/0907.5491} {arXiv:0907.5491 [hep-lat]} \BibitemShut
  {NoStop}%
\bibitem [{\citenamefont {L\"uscher}(2010)}]{Luscher:2010iy}%
  \BibitemOpen
  \bibfield  {author} {\bibinfo {author} {\bibfnamefont {M.}~\bibnamefont
  {L\"uscher}},\ }\href {\doibase 10.1007/JHEP08(2010)071} {\bibfield
  {journal} {\bibinfo  {journal} {JHEP}\ }\textbf {\bibinfo {volume} {08}},\
  \bibinfo {pages} {071} (\bibinfo {year} {2010})},\ \bibinfo {note} {[Erratum:
  JHEP 03, 092 (2014)]},\ \Eprint {http://arxiv.org/abs/1006.4518}
  {arXiv:1006.4518 [hep-lat]} \BibitemShut {NoStop}%
\bibitem [{\citenamefont {L\"uscher}\ and\ \citenamefont
  {Weisz}(2011)}]{Luscher:2011bx}%
  \BibitemOpen
  \bibfield  {author} {\bibinfo {author} {\bibfnamefont {M.}~\bibnamefont
  {L\"uscher}}\ and\ \bibinfo {author} {\bibfnamefont {P.}~\bibnamefont
  {Weisz}},\ }\href {\doibase 10.1007/JHEP02(2011)051} {\bibfield  {journal}
  {\bibinfo  {journal} {JHEP}\ }\textbf {\bibinfo {volume} {02}},\ \bibinfo
  {pages} {051} (\bibinfo {year} {2011})},\ \Eprint
  {http://arxiv.org/abs/1101.0963} {arXiv:1101.0963 [hep-th]} \BibitemShut
  {NoStop}%
%%CITATION = ARXIV:1101.0963;%%
\bibitem [{\citenamefont {Altenkort}\ \emph {et~al.}(2021)\citenamefont
  {Altenkort}, \citenamefont {Eller}, \citenamefont {Kaczmarek}, \citenamefont
  {Mazur}, \citenamefont {Moore},\ and\ \citenamefont
  {Shu}}]{Altenkort:2020fgs}%
  \BibitemOpen
  \bibfield  {author} {\bibinfo {author} {\bibfnamefont {L.}~\bibnamefont
  {Altenkort}}, \bibinfo {author} {\bibfnamefont {A.~M.}\ \bibnamefont
  {Eller}}, \bibinfo {author} {\bibfnamefont {O.}~\bibnamefont {Kaczmarek}},
  \bibinfo {author} {\bibfnamefont {L.}~\bibnamefont {Mazur}}, \bibinfo
  {author} {\bibfnamefont {G.~D.}\ \bibnamefont {Moore}}, \ and\ \bibinfo
  {author} {\bibfnamefont {H.-T.}\ \bibnamefont {Shu}},\ }\href {\doibase
  10.1103/PhysRevD.103.014511} {\bibfield  {journal} {\bibinfo  {journal}
  {Phys. Rev. D}\ }\textbf {\bibinfo {volume} {103}},\ \bibinfo {pages}
  {014511} (\bibinfo {year} {2021})},\ \Eprint
  {http://arxiv.org/abs/2009.13553} {arXiv:2009.13553 [hep-lat]} \BibitemShut
  {NoStop}%
\bibitem [{\citenamefont {Brambilla}\ \emph {et~al.}(2023)\citenamefont
  {Brambilla}, \citenamefont {Leino}, \citenamefont {Mayer-Steudte},\ and\
  \citenamefont {Petreczky}}]{Brambilla:2022xbd}%
  \BibitemOpen
  \bibfield  {author} {\bibinfo {author} {\bibfnamefont {N.}~\bibnamefont
  {Brambilla}}, \bibinfo {author} {\bibfnamefont {V.}~\bibnamefont {Leino}},
  \bibinfo {author} {\bibfnamefont {J.}~\bibnamefont {Mayer-Steudte}}, \ and\
  \bibinfo {author} {\bibfnamefont {P.}~\bibnamefont {Petreczky}} (\bibinfo
  {collaboration} {TUMQCD}),\ }\href {\doibase 10.1103/PhysRevD.107.054508}
  {\bibfield  {journal} {\bibinfo  {journal} {Phys. Rev. D}\ }\textbf {\bibinfo
  {volume} {107}},\ \bibinfo {pages} {054508} (\bibinfo {year} {2023})},\
  \Eprint {http://arxiv.org/abs/2206.02861} {arXiv:2206.02861 [hep-lat]}
  \BibitemShut {NoStop}%
\bibitem [{\citenamefont {Eller}\ and\ \citenamefont
  {Moore}(2018)}]{Eller:2018yje}%
  \BibitemOpen
  \bibfield  {author} {\bibinfo {author} {\bibfnamefont {A.~M.}\ \bibnamefont
  {Eller}}\ and\ \bibinfo {author} {\bibfnamefont {G.~D.}\ \bibnamefont
  {Moore}},\ }\href {\doibase 10.1103/PhysRevD.97.114507} {\bibfield  {journal}
  {\bibinfo  {journal} {Phys. Rev. D}\ }\textbf {\bibinfo {volume} {97}},\
  \bibinfo {pages} {114507} (\bibinfo {year} {2018})},\ \Eprint
  {http://arxiv.org/abs/1802.04562} {arXiv:1802.04562 [hep-lat]} \BibitemShut
  {NoStop}%
\bibitem [{\citenamefont {Eichten}\ and\ \citenamefont
  {Hill}(1990)}]{Eichten:1990vp}%
  \BibitemOpen
  \bibfield  {author} {\bibinfo {author} {\bibfnamefont {E.}~\bibnamefont
  {Eichten}}\ and\ \bibinfo {author} {\bibfnamefont {B.~R.}\ \bibnamefont
  {Hill}},\ }\href {\doibase 10.1016/0370-2693(90)91408-4} {\bibfield
  {journal} {\bibinfo  {journal} {Phys. Lett. B}\ }\textbf {\bibinfo {volume}
  {243}},\ \bibinfo {pages} {427} (\bibinfo {year} {1990})}\BibitemShut
  {NoStop}%
\bibitem [{\citenamefont {Cruz}\ and\ \citenamefont
  {Moore}()}]{guy-david-matching}%
  \BibitemOpen
  \bibfield  {author} {\bibinfo {author} {\bibfnamefont {D.~d.~l.}\
  \bibnamefont {Cruz}}\ and\ \bibinfo {author} {\bibfnamefont {G.~D.}\
  \bibnamefont {Moore}},\ }\href@noop {} {\bibinfo  {journal} {in preparation}\
  }\BibitemShut {NoStop}%
\bibitem [{\citenamefont {Sambataro}\ \emph {et~al.}(2023)\citenamefont
  {Sambataro}, \citenamefont {Minissale}, \citenamefont {Plumari},\ and\
  \citenamefont {Greco}}]{Sambataro:2023tlv}%
  \BibitemOpen
\bibfield  {journal} {  }\bibfield  {author} {\bibinfo {author} {\bibfnamefont
  {M.~L.}\ \bibnamefont {Sambataro}}, \bibinfo {author} {\bibfnamefont
  {V.}~\bibnamefont {Minissale}}, \bibinfo {author} {\bibfnamefont
  {S.}~\bibnamefont {Plumari}}, \ and\ \bibinfo {author} {\bibfnamefont
  {V.}~\bibnamefont {Greco}},\ }\href@noop {} {\  (\bibinfo {year} {2023})},\
  \Eprint {http://arxiv.org/abs/2304.02953} {arXiv:2304.02953 [hep-ph]}
  \BibitemShut {NoStop}%
\bibitem [{\citenamefont {Liu}\ and\ \citenamefont {Rapp}(2020)}]{Liu:2016ysz}%
  \BibitemOpen
  \bibfield  {author} {\bibinfo {author} {\bibfnamefont {S.~Y.~F.}\
  \bibnamefont {Liu}}\ and\ \bibinfo {author} {\bibfnamefont {R.}~\bibnamefont
  {Rapp}},\ }\href {\doibase 10.1140/epja/s10050-020-00024-z} {\bibfield
  {journal} {\bibinfo  {journal} {Eur. Phys. J. A}\ }\textbf {\bibinfo {volume}
  {56}},\ \bibinfo {pages} {44} (\bibinfo {year} {2020})},\ \Eprint
  {http://arxiv.org/abs/1612.09138} {arXiv:1612.09138 [nucl-th]} \BibitemShut
  {NoStop}%
\bibitem [{\citenamefont {Tang}\ \emph {et~al.}(2023)\citenamefont {Tang},
  \citenamefont {Mukherjee}, \citenamefont {Petreczky},\ and\ \citenamefont
  {Rapp}}]{ZhanduoTang:2023ewm}%
  \BibitemOpen
  \bibfield  {author} {\bibinfo {author} {\bibfnamefont {Z.}~\bibnamefont
  {Tang}}, \bibinfo {author} {\bibfnamefont {S.}~\bibnamefont {Mukherjee}},
  \bibinfo {author} {\bibfnamefont {P.}~\bibnamefont {Petreczky}}, \ and\
  \bibinfo {author} {\bibfnamefont {R.}~\bibnamefont {Rapp}},\ }\href@noop {}
  {\  (\bibinfo {year} {2023})},\ \Eprint {http://arxiv.org/abs/2310.18864}
  {arXiv:2310.18864 [hep-lat]} \BibitemShut {NoStop}%
\bibitem [{\citenamefont {Xu}\ \emph {et~al.}(2018)\citenamefont {Xu},
  \citenamefont {Bernhard}, \citenamefont {Bass}, \citenamefont {Nahrgang},\
  and\ \citenamefont {Cao}}]{Xu:2017obm}%
  \BibitemOpen
  \bibfield  {author} {\bibinfo {author} {\bibfnamefont {Y.}~\bibnamefont
  {Xu}}, \bibinfo {author} {\bibfnamefont {J.~E.}\ \bibnamefont {Bernhard}},
  \bibinfo {author} {\bibfnamefont {S.~A.}\ \bibnamefont {Bass}}, \bibinfo
  {author} {\bibfnamefont {M.}~\bibnamefont {Nahrgang}}, \ and\ \bibinfo
  {author} {\bibfnamefont {S.}~\bibnamefont {Cao}},\ }\href {\doibase
  10.1103/PhysRevC.97.014907} {\bibfield  {journal} {\bibinfo  {journal} {Phys.
  Rev. C}\ }\textbf {\bibinfo {volume} {97}},\ \bibinfo {pages} {014907}
  (\bibinfo {year} {2018})},\ \Eprint {http://arxiv.org/abs/1710.00807}
  {arXiv:1710.00807 [nucl-th]} \BibitemShut {NoStop}%
\bibitem [{\citenamefont {Acharya}\ \emph {et~al.}(2022)\citenamefont {Acharya}
  \emph {et~al.}}]{ALICE:2021rxa}%
  \BibitemOpen
  \bibfield  {author} {\bibinfo {author} {\bibfnamefont {S.}~\bibnamefont
  {Acharya}} \emph {et~al.} (\bibinfo {collaboration} {ALICE}),\ }\href
  {\doibase 10.1007/JHEP01(2022)174} {\bibfield  {journal} {\bibinfo  {journal}
  {JHEP}\ }\textbf {\bibinfo {volume} {01}},\ \bibinfo {pages} {174} (\bibinfo
  {year} {2022})},\ \Eprint {http://arxiv.org/abs/2110.09420} {arXiv:2110.09420
  [nucl-ex]} \BibitemShut {NoStop}%
\bibitem [{\citenamefont {Caron-Huot}\ and\ \citenamefont
  {Moore}(2008)}]{CaronHuot:2007gq}%
  \BibitemOpen
  \bibfield  {author} {\bibinfo {author} {\bibfnamefont {S.}~\bibnamefont
  {Caron-Huot}}\ and\ \bibinfo {author} {\bibfnamefont {G.~D.}\ \bibnamefont
  {Moore}},\ }\href {\doibase 10.1103/PhysRevLett.100.052301} {\bibfield
  {journal} {\bibinfo  {journal} {Phys. Rev. Lett.}\ }\textbf {\bibinfo
  {volume} {100}},\ \bibinfo {pages} {052301} (\bibinfo {year} {2008})},\
  \Eprint {http://arxiv.org/abs/0708.4232} {arXiv:0708.4232 [hep-ph]}
  \BibitemShut {NoStop}%
\bibitem [{\citenamefont {Eller}(2021)}]{Eller:2021qpp}%
  \BibitemOpen
  \bibfield  {author} {\bibinfo {author} {\bibfnamefont {A.~M.}\ \bibnamefont
  {Eller}},\ }\emph {\bibinfo {title} {{The Color-Electric Field Correlator
  under Gradient Flow at next-to-leading Order in Quantum Chromodynamics}}},\
  \href {\doibase 10.26083/tuprints-00017610} {Ph.D. thesis},\ \bibinfo
  {school} {Technische Universit\"at Darmstadt} (\bibinfo {year}
  {2021})\BibitemShut {NoStop}%
\bibitem [{\citenamefont {Banerjee}\ \emph
  {et~al.}(2022{\natexlab{a}})\citenamefont {Banerjee}, \citenamefont {Datta},\
  and\ \citenamefont {Laine}}]{Banerjee:2022uge}%
  \BibitemOpen
  \bibfield  {author} {\bibinfo {author} {\bibfnamefont {D.}~\bibnamefont
  {Banerjee}}, \bibinfo {author} {\bibfnamefont {S.}~\bibnamefont {Datta}}, \
  and\ \bibinfo {author} {\bibfnamefont {M.}~\bibnamefont {Laine}},\ }\href
  {\doibase 10.1007/JHEP08(2022)128} {\bibfield  {journal} {\bibinfo  {journal}
  {JHEP}\ }\textbf {\bibinfo {volume} {08}},\ \bibinfo {pages} {128} (\bibinfo
  {year} {2022}{\natexlab{a}})},\ \Eprint {http://arxiv.org/abs/2204.14075}
  {arXiv:2204.14075 [hep-lat]} \BibitemShut {NoStop}%
\bibitem [{\citenamefont {Laine}(2021)}]{Laine:2021uzs}%
  \BibitemOpen
  \bibfield  {author} {\bibinfo {author} {\bibfnamefont {M.}~\bibnamefont
  {Laine}},\ }\href {\doibase 10.1007/JHEP06(2021)139} {\bibfield  {journal}
  {\bibinfo  {journal} {JHEP}\ }\textbf {\bibinfo {volume} {06}},\ \bibinfo
  {pages} {139} (\bibinfo {year} {2021})},\ \Eprint
  {http://arxiv.org/abs/2103.14270} {arXiv:2103.14270 [hep-ph]} \BibitemShut
  {NoStop}%
\bibitem [{\citenamefont {Francis}\ \emph {et~al.}(2015)\citenamefont
  {Francis}, \citenamefont {Kaczmarek}, \citenamefont {Laine}, \citenamefont
  {Neuhaus},\ and\ \citenamefont {Ohno}}]{Francis:2015daa}%
  \BibitemOpen
  \bibfield  {author} {\bibinfo {author} {\bibfnamefont {A.}~\bibnamefont
  {Francis}}, \bibinfo {author} {\bibfnamefont {O.}~\bibnamefont {Kaczmarek}},
  \bibinfo {author} {\bibfnamefont {M.}~\bibnamefont {Laine}}, \bibinfo
  {author} {\bibfnamefont {T.}~\bibnamefont {Neuhaus}}, \ and\ \bibinfo
  {author} {\bibfnamefont {H.}~\bibnamefont {Ohno}},\ }\href {\doibase
  10.1103/PhysRevD.92.116003} {\bibfield  {journal} {\bibinfo  {journal} {Phys.
  Rev. D}\ }\textbf {\bibinfo {volume} {92}},\ \bibinfo {pages} {116003}
  (\bibinfo {year} {2015})},\ \Eprint {http://arxiv.org/abs/1508.04543}
  {arXiv:1508.04543 [hep-lat]} \BibitemShut {NoStop}%
\bibitem [{\citenamefont {Brambilla}\ \emph {et~al.}(2020)\citenamefont
  {Brambilla}, \citenamefont {Leino}, \citenamefont {Petreczky},\ and\
  \citenamefont {Vairo}}]{Brambilla:2020siz}%
  \BibitemOpen
  \bibfield  {author} {\bibinfo {author} {\bibfnamefont {N.}~\bibnamefont
  {Brambilla}}, \bibinfo {author} {\bibfnamefont {V.}~\bibnamefont {Leino}},
  \bibinfo {author} {\bibfnamefont {P.}~\bibnamefont {Petreczky}}, \ and\
  \bibinfo {author} {\bibfnamefont {A.}~\bibnamefont {Vairo}},\ }\href
  {\doibase 10.1103/PhysRevD.102.074503} {\bibfield  {journal} {\bibinfo
  {journal} {Phys. Rev. D}\ }\textbf {\bibinfo {volume} {102}},\ \bibinfo
  {pages} {074503} (\bibinfo {year} {2020})},\ \Eprint
  {http://arxiv.org/abs/2007.10078} {arXiv:2007.10078 [hep-lat]} \BibitemShut
  {NoStop}%
\bibitem [{\citenamefont {Kajantie}\ \emph {et~al.}(1997)\citenamefont
  {Kajantie}, \citenamefont {Laine}, \citenamefont {Rummukainen},\ and\
  \citenamefont {Shaposhnikov}}]{Kajantie:1997tt}%
  \BibitemOpen
  \bibfield  {author} {\bibinfo {author} {\bibfnamefont {K.}~\bibnamefont
  {Kajantie}}, \bibinfo {author} {\bibfnamefont {M.}~\bibnamefont {Laine}},
  \bibinfo {author} {\bibfnamefont {K.}~\bibnamefont {Rummukainen}}, \ and\
  \bibinfo {author} {\bibfnamefont {M.~E.}\ \bibnamefont {Shaposhnikov}},\
  }\href {\doibase 10.1016/S0550-3213(97)00425-2} {\bibfield  {journal}
  {\bibinfo  {journal} {Nucl. Phys. B}\ }\textbf {\bibinfo {volume} {503}},\
  \bibinfo {pages} {357} (\bibinfo {year} {1997})},\ \Eprint
  {http://arxiv.org/abs/hep-ph/9704416} {arXiv:hep-ph/9704416} \BibitemShut
  {NoStop}%
\bibitem [{\citenamefont {Burnier}\ and\ \citenamefont
  {Laine}(2012)}]{Burnier:2012ze}%
  \BibitemOpen
  \bibfield  {author} {\bibinfo {author} {\bibfnamefont {Y.}~\bibnamefont
  {Burnier}}\ and\ \bibinfo {author} {\bibfnamefont {M.}~\bibnamefont
  {Laine}},\ }\href {\doibase 10.1007/JHEP11(2012)086} {\bibfield  {journal}
  {\bibinfo  {journal} {JHEP}\ }\textbf {\bibinfo {volume} {11}},\ \bibinfo
  {pages} {086} (\bibinfo {year} {2012})},\ \Eprint
  {http://arxiv.org/abs/1210.1064} {arXiv:1210.1064 [hep-ph]} \BibitemShut
  {NoStop}%
\bibitem [{\citenamefont {Petreczky}(2009)}]{Petreczky:2008px}%
  \BibitemOpen
  \bibfield  {author} {\bibinfo {author} {\bibfnamefont {P.}~\bibnamefont
  {Petreczky}},\ }\href {\doibase 10.1140/epjc/s10052-009-0942-1} {\bibfield
  {journal} {\bibinfo  {journal} {Eur. Phys. J. C}\ }\textbf {\bibinfo {volume}
  {62}},\ \bibinfo {pages} {85} (\bibinfo {year} {2009})},\ \Eprint
  {http://arxiv.org/abs/0810.0258} {arXiv:0810.0258 [hep-lat]} \BibitemShut
  {NoStop}%
\bibitem [{\citenamefont {Bellwied}\ \emph {et~al.}(2015)\citenamefont
  {Bellwied}, \citenamefont {Borsanyi}, \citenamefont {Fodor}, \citenamefont
  {Katz}, \citenamefont {Pasztor}, \citenamefont {Ratti},\ and\ \citenamefont
  {Szabo}}]{Bellwied:2015lba}%
  \BibitemOpen
  \bibfield  {author} {\bibinfo {author} {\bibfnamefont {R.}~\bibnamefont
  {Bellwied}}, \bibinfo {author} {\bibfnamefont {S.}~\bibnamefont {Borsanyi}},
  \bibinfo {author} {\bibfnamefont {Z.}~\bibnamefont {Fodor}}, \bibinfo
  {author} {\bibfnamefont {S.~D.}\ \bibnamefont {Katz}}, \bibinfo {author}
  {\bibfnamefont {A.}~\bibnamefont {Pasztor}}, \bibinfo {author} {\bibfnamefont
  {C.}~\bibnamefont {Ratti}}, \ and\ \bibinfo {author} {\bibfnamefont {K.~K.}\
  \bibnamefont {Szabo}},\ }\href {\doibase 10.1103/PhysRevD.92.114505}
  {\bibfield  {journal} {\bibinfo  {journal} {Phys. Rev. D}\ }\textbf {\bibinfo
  {volume} {92}},\ \bibinfo {pages} {114505} (\bibinfo {year} {2015})},\
  \Eprint {http://arxiv.org/abs/1507.04627} {arXiv:1507.04627 [hep-lat]}
  \BibitemShut {NoStop}%
\bibitem [{\citenamefont {Mazur}(2021)}]{Mazur:2021zgi}%
  \BibitemOpen
  \bibfield  {author} {\bibinfo {author} {\bibfnamefont {L.}~\bibnamefont
  {Mazur}},\ }\emph {\bibinfo {title} {{Topological Aspects in Lattice QCD}}},\
  \href {\doibase 10.4119/unibi/2956493} {Ph.D. thesis},\ \bibinfo  {school}
  {Bielefeld U.} (\bibinfo {year} {2021})\BibitemShut {NoStop}%
\bibitem [{\citenamefont {Bollweg}\ \emph {et~al.}(2022)\citenamefont
  {Bollweg}, \citenamefont {Altenkort}, \citenamefont {Clarke}, \citenamefont
  {Kaczmarek}, \citenamefont {Mazur}, \citenamefont {Schmidt}, \citenamefont
  {Scior},\ and\ \citenamefont {Shu}}]{Bollweg:2021cvl}%
  \BibitemOpen
  \bibfield  {author} {\bibinfo {author} {\bibfnamefont {D.}~\bibnamefont
  {Bollweg}}, \bibinfo {author} {\bibfnamefont {L.}~\bibnamefont {Altenkort}},
  \bibinfo {author} {\bibfnamefont {D.~A.}\ \bibnamefont {Clarke}}, \bibinfo
  {author} {\bibfnamefont {O.}~\bibnamefont {Kaczmarek}}, \bibinfo {author}
  {\bibfnamefont {L.}~\bibnamefont {Mazur}}, \bibinfo {author} {\bibfnamefont
  {C.}~\bibnamefont {Schmidt}}, \bibinfo {author} {\bibfnamefont
  {P.}~\bibnamefont {Scior}}, \ and\ \bibinfo {author} {\bibfnamefont {H.-T.}\
  \bibnamefont {Shu}},\ }\href {\doibase 10.22323/1.396.0196} {\bibfield
  {journal} {\bibinfo  {journal} {PoS}\ }\textbf {\bibinfo {volume}
  {LATTICE2021}},\ \bibinfo {pages} {196} (\bibinfo {year} {2022})},\ \Eprint
  {http://arxiv.org/abs/2111.10354} {arXiv:2111.10354 [hep-lat]} \BibitemShut
  {NoStop}%
\bibitem [{\citenamefont {Mazur}\ \emph {et~al.}(2023)\citenamefont {Mazur}
  \emph {et~al.}}]{HotQCD:2023ghu}%
  \BibitemOpen
  \bibfield  {author} {\bibinfo {author} {\bibfnamefont {L.}~\bibnamefont
  {Mazur}} \emph {et~al.} (\bibinfo {collaboration} {HotQCD}),\ }\href@noop {}
  {\  (\bibinfo {year} {2023})},\ \Eprint {http://arxiv.org/abs/2306.01098}
  {arXiv:2306.01098 [hep-lat]} \BibitemShut {NoStop}%
\bibitem [{\citenamefont {Altenkort}\ \emph
  {et~al.}(2023{\natexlab{b}})\citenamefont {Altenkort}, \citenamefont {de~la
  Cruz}, \citenamefont {Kaczmarek}, \citenamefont {R.}, \citenamefont {G.D.},
  \citenamefont {S.}, \citenamefont {P.}, \citenamefont {Shu},\ and\
  \citenamefont {Stendebach}}]{datapublication}%
  \BibitemOpen
  \bibfield  {author} {\bibinfo {author} {\bibfnamefont {L.}~\bibnamefont
  {Altenkort}}, \bibinfo {author} {\bibfnamefont {D.}~\bibnamefont {de~la
  Cruz}}, \bibinfo {author} {\bibfnamefont {O.}~\bibnamefont {Kaczmarek}},
  \bibinfo {author} {\bibfnamefont {L.}~\bibnamefont {R.}}, \bibinfo {author}
  {\bibfnamefont {M.}~\bibnamefont {G.D.}}, \bibinfo {author} {\bibfnamefont
  {M.}~\bibnamefont {S.}}, \bibinfo {author} {\bibfnamefont {P.}~\bibnamefont
  {P.}}, \bibinfo {author} {\bibfnamefont {H.-T.}\ \bibnamefont {Shu}}, \ and\
  \bibinfo {author} {\bibfnamefont {S.}~\bibnamefont {Stendebach}},\ }\href
  {\doibase 10.4119/unibi/2985523} {\  (\bibinfo {year} {2023}{\natexlab{b}}),\
  10.4119/unibi/2985523}\BibitemShut {NoStop}%
\bibitem [{\citenamefont {Clark}\ and\ \citenamefont
  {Kennedy}(2004)}]{Clark:2003na}%
  \BibitemOpen
  \bibfield  {author} {\bibinfo {author} {\bibfnamefont {M.~A.}\ \bibnamefont
  {Clark}}\ and\ \bibinfo {author} {\bibfnamefont {A.~D.}\ \bibnamefont
  {Kennedy}},\ }\href {\doibase 10.1016/S0920-5632(03)02732-4} {\bibfield
  {journal} {\bibinfo  {journal} {Nucl. Phys. B Proc. Suppl.}\ }\textbf
  {\bibinfo {volume} {129}},\ \bibinfo {pages} {850} (\bibinfo {year}
  {2004})},\ \Eprint {http://arxiv.org/abs/hep-lat/0309084}
  {arXiv:hep-lat/0309084} \BibitemShut {NoStop}%
\bibitem [{\citenamefont {Clark}\ \emph {et~al.}(2005)\citenamefont {Clark},
  \citenamefont {Kennedy},\ and\ \citenamefont {Sroczynski}}]{Clark:2004cp}%
  \BibitemOpen
  \bibfield  {author} {\bibinfo {author} {\bibfnamefont {M.~A.}\ \bibnamefont
  {Clark}}, \bibinfo {author} {\bibfnamefont {A.~D.}\ \bibnamefont {Kennedy}},
  \ and\ \bibinfo {author} {\bibfnamefont {Z.}~\bibnamefont {Sroczynski}},\
  }\href {\doibase 10.1016/j.nuclphysbps.2004.11.192} {\bibfield  {journal}
  {\bibinfo  {journal} {Nucl. Phys. B Proc. Suppl.}\ }\textbf {\bibinfo
  {volume} {140}},\ \bibinfo {pages} {835} (\bibinfo {year} {2005})},\ \Eprint
  {http://arxiv.org/abs/hep-lat/0409133} {arXiv:hep-lat/0409133} \BibitemShut
  {NoStop}%
\bibitem [{\citenamefont {Bazavov}\ \emph
  {et~al.}(2014{\natexlab{a}})\citenamefont {Bazavov} \emph
  {et~al.}}]{HotQCD:2014kol}%
  \BibitemOpen
  \bibfield  {author} {\bibinfo {author} {\bibfnamefont {A.}~\bibnamefont
  {Bazavov}} \emph {et~al.} (\bibinfo {collaboration} {HotQCD}),\ }\href
  {\doibase 10.1103/PhysRevD.90.094503} {\bibfield  {journal} {\bibinfo
  {journal} {Phys. Rev. D}\ }\textbf {\bibinfo {volume} {90}},\ \bibinfo
  {pages} {094503} (\bibinfo {year} {2014}{\natexlab{a}})},\ \Eprint
  {http://arxiv.org/abs/1407.6387} {arXiv:1407.6387 [hep-lat]} \BibitemShut
  {NoStop}%
\bibitem [{\citenamefont {Bazavov}\ \emph {et~al.}(2018)\citenamefont
  {Bazavov}, \citenamefont {Petreczky},\ and\ \citenamefont
  {Weber}}]{Bazavov:2017dsy}%
  \BibitemOpen
  \bibfield  {author} {\bibinfo {author} {\bibfnamefont {A.}~\bibnamefont
  {Bazavov}}, \bibinfo {author} {\bibfnamefont {P.}~\bibnamefont {Petreczky}},
  \ and\ \bibinfo {author} {\bibfnamefont {J.}~\bibnamefont {Weber}},\ }\href
  {\doibase 10.1103/PhysRevD.97.014510} {\bibfield  {journal} {\bibinfo
  {journal} {Phys. Rev. D}\ }\textbf {\bibinfo {volume} {97}},\ \bibinfo
  {pages} {014510} (\bibinfo {year} {2018})},\ \Eprint
  {http://arxiv.org/abs/1710.05024} {arXiv:1710.05024 [hep-lat]} \BibitemShut
  {NoStop}%
\bibitem [{\citenamefont {Bazavov}\ \emph {et~al.}(2010)\citenamefont {Bazavov}
  \emph {et~al.}}]{MILC:2010hzw}%
  \BibitemOpen
  \bibfield  {author} {\bibinfo {author} {\bibfnamefont {A.}~\bibnamefont
  {Bazavov}} \emph {et~al.} (\bibinfo {collaboration} {MILC}),\ }\href
  {\doibase 10.22323/1.105.0074} {\bibfield  {journal} {\bibinfo  {journal}
  {PoS}\ }\textbf {\bibinfo {volume} {LATTICE2010}},\ \bibinfo {pages} {074}
  (\bibinfo {year} {2010})},\ \Eprint {http://arxiv.org/abs/1012.0868}
  {arXiv:1012.0868 [hep-lat]} \BibitemShut {NoStop}%
\bibitem [{\citenamefont {Bazavov}\ and\ \citenamefont
  {Petreczky}(2010)}]{Bazavov:2010sb}%
  \BibitemOpen
  \bibfield  {author} {\bibinfo {author} {\bibfnamefont {A.}~\bibnamefont
  {Bazavov}}\ and\ \bibinfo {author} {\bibfnamefont {P.}~\bibnamefont
  {Petreczky}} (\bibinfo {collaboration} {HotQCD}),\ }\href {\doibase
  10.1088/1742-6596/230/1/012014} {\bibfield  {journal} {\bibinfo  {journal}
  {J. Phys. Conf. Ser.}\ }\textbf {\bibinfo {volume} {230}},\ \bibinfo {pages}
  {012014} (\bibinfo {year} {2010})},\ \Eprint {http://arxiv.org/abs/1005.1131}
  {arXiv:1005.1131 [hep-lat]} \BibitemShut {NoStop}%
\bibitem [{\citenamefont {Herren}\ and\ \citenamefont
  {Steinhauser}(2018)}]{Herren:2017osy}%
  \BibitemOpen
  \bibfield  {author} {\bibinfo {author} {\bibfnamefont {F.}~\bibnamefont
  {Herren}}\ and\ \bibinfo {author} {\bibfnamefont {M.}~\bibnamefont
  {Steinhauser}},\ }\href {\doibase 10.1016/j.cpc.2017.11.014} {\bibfield
  {journal} {\bibinfo  {journal} {Comput. Phys. Commun.}\ }\textbf {\bibinfo
  {volume} {224}},\ \bibinfo {pages} {333} (\bibinfo {year} {2018})},\ \Eprint
  {http://arxiv.org/abs/1703.03751} {arXiv:1703.03751 [hep-ph]} \BibitemShut
  {NoStop}%
\bibitem [{\citenamefont {Chetyrkin}\ \emph {et~al.}(2000)\citenamefont
  {Chetyrkin}, \citenamefont {Kuhn},\ and\ \citenamefont
  {Steinhauser}}]{Chetyrkin:2000yt}%
  \BibitemOpen
  \bibfield  {author} {\bibinfo {author} {\bibfnamefont {K.~G.}\ \bibnamefont
  {Chetyrkin}}, \bibinfo {author} {\bibfnamefont {J.~H.}\ \bibnamefont {Kuhn}},
  \ and\ \bibinfo {author} {\bibfnamefont {M.}~\bibnamefont {Steinhauser}},\
  }\href {\doibase 10.1016/S0010-4655(00)00155-7} {\bibfield  {journal}
  {\bibinfo  {journal} {Comput. Phys. Commun.}\ }\textbf {\bibinfo {volume}
  {133}},\ \bibinfo {pages} {43} (\bibinfo {year} {2000})},\ \Eprint
  {http://arxiv.org/abs/hep-ph/0004189} {arXiv:hep-ph/0004189} \BibitemShut
  {NoStop}%
\bibitem [{\citenamefont {Aoki}\ \emph {et~al.}(2022)\citenamefont {Aoki} \emph
  {et~al.}}]{FlavourLatticeAveragingGroupFLAG:2021npn}%
  \BibitemOpen
  \bibfield  {author} {\bibinfo {author} {\bibfnamefont {Y.}~\bibnamefont
  {Aoki}} \emph {et~al.} (\bibinfo {collaboration} {Flavour Lattice Averaging
  Group (FLAG)}),\ }\href {\doibase 10.1140/epjc/s10052-022-10536-1} {\bibfield
   {journal} {\bibinfo  {journal} {Eur. Phys. J. C}\ }\textbf {\bibinfo
  {volume} {82}},\ \bibinfo {pages} {869} (\bibinfo {year} {2022})},\ \Eprint
  {http://arxiv.org/abs/2111.09849} {arXiv:2111.09849 [hep-lat]} \BibitemShut
  {NoStop}%
\bibitem [{\citenamefont {Bazavov}\ \emph
  {et~al.}(2014{\natexlab{b}})\citenamefont {Bazavov} \emph
  {et~al.}}]{Bazavov:2014pvz}%
  \BibitemOpen
  \bibfield  {author} {\bibinfo {author} {\bibfnamefont {A.}~\bibnamefont
  {Bazavov}} \emph {et~al.} (\bibinfo {collaboration} {HotQCD}),\ }\href
  {\doibase 10.1103/PhysRevD.90.094503} {\bibfield  {journal} {\bibinfo
  {journal} {Phys. Rev.}\ }\textbf {\bibinfo {volume} {D90}},\ \bibinfo {pages}
  {094503} (\bibinfo {year} {2014}{\natexlab{b}})},\ \Eprint
  {http://arxiv.org/abs/1407.6387} {arXiv:1407.6387 [hep-lat]} \BibitemShut
  {NoStop}%
%%CITATION = ARXIV:1407.6387;%%
\bibitem [{\citenamefont {Stendebach}(2022)}]{2022-Stendebach}%
  \BibitemOpen
  \bibfield  {author} {\bibinfo {author} {\bibfnamefont {S.}~\bibnamefont
  {Stendebach}},\ }\emph {\bibinfo {title} {{Perturbative analysis of operators
  under improved gradient flow in lattice QCD}}},\ \href {\doibase
  10.26083/tuprints-00023185} {Master's thesis},\ \bibinfo  {school}
  {Technische Universität Darmstadt} (\bibinfo {year} {2022})\BibitemShut
  {NoStop}%
\bibitem [{\citenamefont {Banerjee}\ \emph
  {et~al.}(2022{\natexlab{b}})\citenamefont {Banerjee}, \citenamefont {Gavai},
  \citenamefont {Datta},\ and\ \citenamefont {Majumdar}}]{Banerjee:2022gen}%
  \BibitemOpen
  \bibfield  {author} {\bibinfo {author} {\bibfnamefont {D.}~\bibnamefont
  {Banerjee}}, \bibinfo {author} {\bibfnamefont {R.}~\bibnamefont {Gavai}},
  \bibinfo {author} {\bibfnamefont {S.}~\bibnamefont {Datta}}, \ and\ \bibinfo
  {author} {\bibfnamefont {P.}~\bibnamefont {Majumdar}},\ }\href@noop {} {\
  (\bibinfo {year} {2022}{\natexlab{b}})},\ \Eprint
  {http://arxiv.org/abs/2206.15471} {arXiv:2206.15471 [hep-ph]} \BibitemShut
  {NoStop}%
\end{thebibliography}%
\newpage
\begin{widetext}
\vspace{2em}
\begin{center}
    {\Large \bf Supplemental Materials}
\end{center}
\vspace{2em}

In these supporting materials we provide the details mentioned in the Letter for the calculation of the finite mass correction to the heavy quark momentum diffusion coefficient via $B$-field correlators.
We start with Sec.~\ref{supp-mat:setup} by giving additional details to the lattice setup. 
In Sec.~\ref{supp-mat:coupling} we describe the calculation of the matching factor $\Zmatch$. 
Sec.~\ref{supp-mat:extrap} is devoted to the discussions on the continuum
and flow time extrapolations of the $B$-field correlators. In Sec.
\ref{supp-mat:SPF} we detail the modeling of the spectral function. The obtained $\kappa_B$ is compared to other estimates and $\kappa_E$ in Sec.~\ref{sec:compare-kappa}. We close by checking the dependence of $2\pi TD_s$ on the thermal charm and bottom quark mass in Sec.~\ref{sec:mass-dep}.

\section{Lattice setup}
\label{supp-mat:setup}

\begin{table}[htb]
\centering
\begin{tabular}{|c|cccccc|}
\hline
$T$ [MeV] & $\beta$ & $a m_s$ & $a m_l$ & $N_{\sigma}$ & $N_{\tau}$ & \# conf.\\
\hline
195 & 7.570	& 0.01973 &	0.003946 & 64 & 20 & 5899 \\
    & 7.777	& 0.01601 &	0.003202 & 64 & 24 & 3435 \\
    & 8.249 & 0.01011 & 0.002022 & 96 & 36 & 2256 \\
\hline
220 & 7.704	& 0.01723 &	0.003446 & 64 & 20 & 7923 \\
    & 7.913	& 0.01400 &	0.002800 & 64 & 24 & 2715 \\
    & 8.249 & 0.01011 & 0.002022 & 96 & 32 & 912 \\
\hline 
251 & 7.857	& 0.01479 &	0.002958 & 64 & 20 & 6786 \\
    & 8.068	& 0.01204 &	0.002408 & 64 & 24 & 5325 \\
    & 8.249 & 0.01011 & 0.002022 & 96 & 28 & 1680 \\
\hline
293 & 8.036	& 0.01241 &	0.002482 & 64 & 20 & 6534 \\
    & 8.147	& 0.01115 &	0.002230 & 64 & 22 & 9101 \\
    & 8.249 & 0.01011 & 0.002022 & 96 & 24 & 688 \\
    \hline
352 & 8.249 & 0.01011 & 0.002022 & 96 & 20 & 2488 \\
\hline
\end{tabular}
\caption{The temperatures used in this calculation and the $\beta$ values, bare quark masses, lattice sizes and the number of configurations at each temperature.}
\label{tab:paramc}
\end{table}

The gauge configurations in this study were generated using a Rational Hybrid Monte Carlo (RHMC) algorithm~\cite{Clark:2003na,Clark:2004cp}. 
The configurations are saved after every 10 trajectories with acceptance rate $\approx80\%$. 
The dynamical strange quark mass $m_s$ is at its physical value and the light quarks (up/down) are carrying mass $m_l=m_s/5$, corresponding to the pion mass of $320$ MeV.
The masses are obtained from the parametrization of the lines of constant physics from~\cite{HotQCD:2014kol}. Here we consider four temperatures $T=$195, 220, 251 and 293 MeV.
At each temperature three lattices are used to perform the continuum extrapolation. The finest lattices are of size $96^3\times N_\tau$, where $N_\tau=24,28$ and 36. 
All the finest lattices have the same $\beta=8.249$, corresponding to $1/a=7.036$ GeV determined using the $r_1$-scale \cite{Bazavov:2017dsy}, with $r_1=0.3106\,\mathrm{fm}$ taken from \cite{MILC:2010hzw}. The two coarser lattices are of size $64^3\times N_\tau$, where $N_\tau=$20, 22/24,  generated with a $\beta$ value that gives the desired temperature, again determined through the $r_1$ scale.
The lattice parameters of the ensembles used in this calculation, including the number of gauge configurations are summarized in Tab. \ref{tab:paramc}.  

To account for the possible auto-correlation residing in the correlators we estimate the auto-correlation time by analyzing the Polyakov loop at the maximum flow time allowed in this study $\sqrt{8\tau_\mathrm{F}}T=0.15$. The Polyakov loop, as part of the correlator, exhibits stronger auto-correlation than the correlator itself. Thus its auto-correlation can serve as a conservative estimate for the auto-correlation of the correlator. The auto-correlation also becomes stronger at larger flow time. Therefore our strategy of calculating the auto-correlation is robust. The obtained integrated auto-correlation time is $\sim 10-30$ in terms of the number of configurations. Then we group the configurations of that size to make independent bins that are passed to the bootstrap analysis afterwards. 

When presenting the temperature dependence of the obtained heavy quark momentum diffusion coefficient, it is conventional in the literature to express it in terms of $T/T_c$.
$T_c$ in the quenched case means the confinement/deconfinement transition temperature. In full QCD it makes more sense to associate $T_c$ with the chiral crossover temperature, which can be determined by locating the peak location of the disconnected chiral susceptibility. A modelling of the peak location of the disconnected chiral susceptibility suggests that $T_c=180(2)\,\mathrm{MeV}$ \cite{Bazavov:2010sb} 
for the quark masses used in this study. 

\section{Determination of the matching factor}
\label{supp-mat:coupling}
The matching factor from the $B$-field correlator in the gradient flow scheme to the physical correlator is obtained using a three-step matching procedure. 
We first match the gradient flow scheme to the $\MSBAR$ scheme at some
UV scale $\mubaruv$ \cite{guy-david-matching}. Then evolve the $B$-field operators from scale $\mubaruv$ to some thermal scale $\mubarir$, and finally match to the physical
correlator using the result of Ref. \cite{Laine:2021uzs}.
The matching factor reads \cite{guy-david-matching}
\begin{align}
    \label{eq:Ztotal}
    \ln \Zmatch =  \int^{\mubaruv^2}_{\mubarir^2}\gamma_0 \gsqms(\mubar)
    \frac{d\mubar^2}{\mubar^2} + \gamma_0 \gsqms(\mubarir) \left[
    \ln \frac{\mubarir^2}{(4\pi T)^2} - 2 + 2 \gammaE \right] - \gamma_0 \gsqms(\mubaruv) \left[ \ln\frac{\mubaruv^2}{4\, \muflow^2} + \gammaE \right], 
\end{align}
where $\gamma_0 = 3/8\pi^2$ is the leading-order anomalous dimension of a $B$-field. 
In the above formula we use $\MSBAR$ coupling constant $\gsqms$ that is calculated using the RunDec package \cite{Herren:2017osy,Chetyrkin:2000yt} with $N_f=3$,
$\Lambda_{\MSBAR}$=0.339 GeV~\cite{FlavourLatticeAveragingGroupFLAG:2021npn} at 5-loop. This is justified because at this order the difference between the $\MSBAR$ coupling
constant and gradient flow coupling constant can be neglected.
Different choices of the scales lead to four different $\Zmatch(\mubarir,\mubaruv,\muflow)$, which are shown as a function of $\sqrt{8\tauf}/r_0$ in Fig.~\ref{fig:zmatch}. $r_0=0.469$ fm is taken from \cite{Bazavov:2014pvz}.
Changing $\mubarir$ at fixed $\mubaruv$ amounts to an overall multiplicative shift, while changing $\mubaruv$ changes the shape of the curve.
However, the small-$\tauf$ limiting behavior is not sensitive to $\mubaruv$.
This indicates that the obtained $\kappa_B$ will show a small difference among different $\mubaruv$ choices.

\begin{figure}[tbh]
\centerline{
\includegraphics[width=0.5\textwidth]{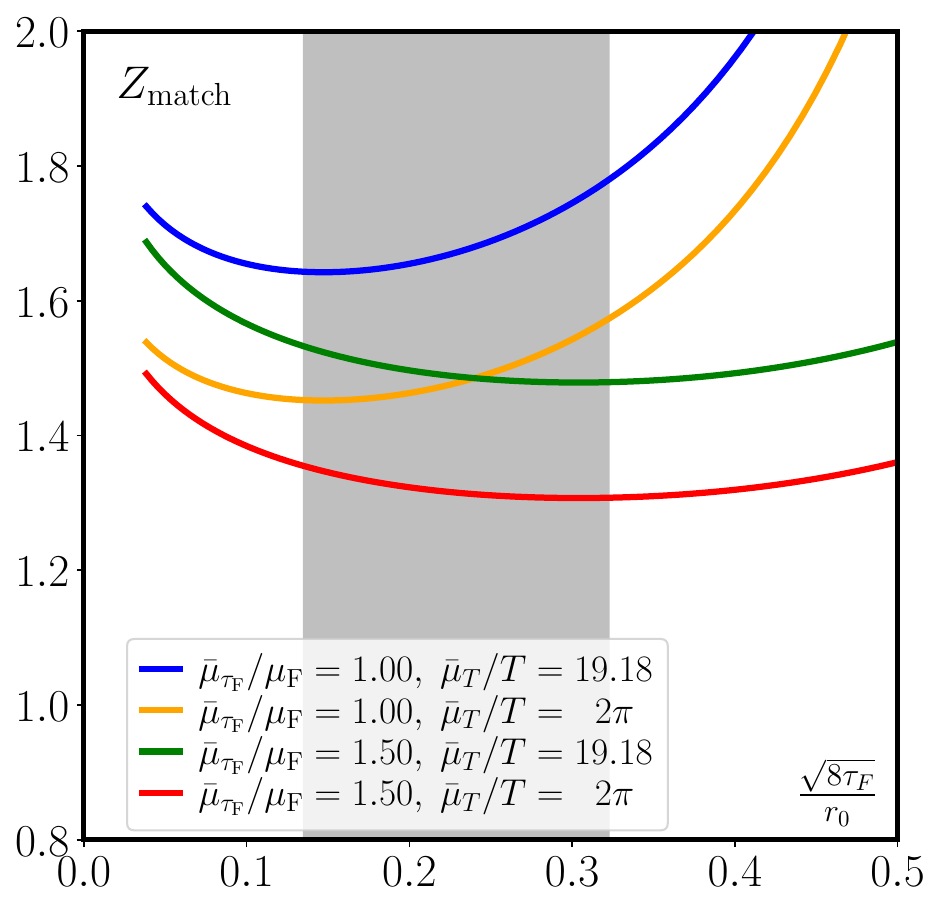}}
\caption{$\Zmatch(\mubarir,\mubaruv,\muflow)$ calculated for different scales at $T=195$ MeV as an example. The grey band covers the flow time range valid for the $\tauf\rightarrow 0$ extrapolation, see Sec.~\ref{supp-mat:extrap}.}
\label{fig:zmatch}
\end{figure}

\section{Continuum and flow time extrapolations}
\label{supp-mat:extrap}

\begin{figure}[tbh]
\centerline{
\includegraphics[width=0.5\textwidth]{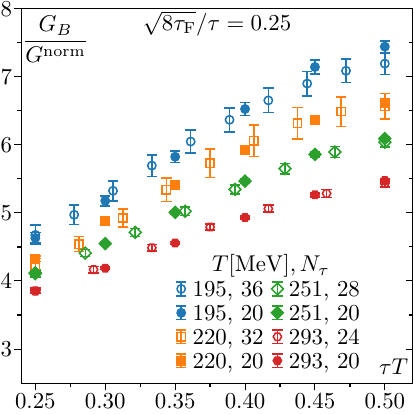}
\includegraphics[width=0.5\textwidth]{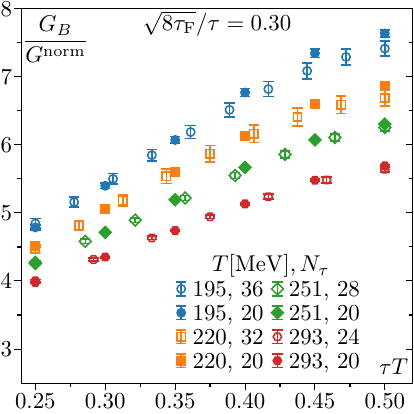}
}
    \caption{The $B$-field correlators in the gradient flow scheme for different temperatures calculated on the finest (open symbols) and coarsest lattices (filled symbols) at relative flow time $\sqrt{8\tauf}/\tau=0.25$ (left) and 0.30 (right). The lattice results have been normalized
    by $\Gnorm$ obtained at non-zero flow time and lattice spacing.}
    \label{fig:corr-before-cont}
\end{figure}

\begin{figure*}[tbh]
\includegraphics[page=21,width=0.46\textwidth]{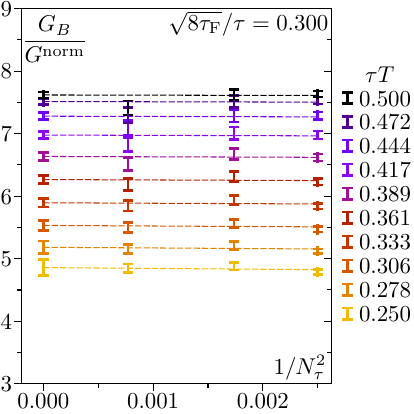}
\includegraphics[page=21,width=0.46\textwidth]{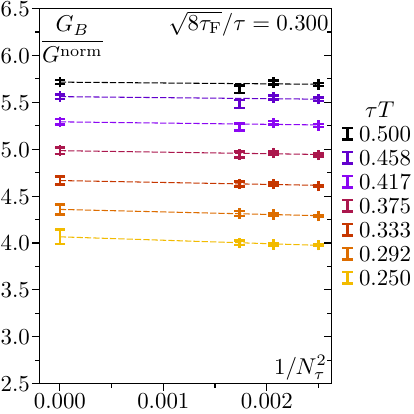}
    \caption{The lattice spacing dependence of the
    $B$-field correlator at 
    $T=195\,\mathrm{MeV}$ (left) and $T=293\,\mathrm{MeV}$ (right) at $\sqrt{8\tauf}/\tau=0.3$. The dashed lines show the central values of the continuum extrapolations.}
    \label{fig:cont_extra_BB}
\end{figure*}
As discussed in the main text, for the analysis of the $B$-field correlator and the continuum extrapolations we need the normalization correlator, $G^{\mathrm{norm}}(\tau,T,N_\tau,\tauf)=G_B^{\mathrm{LO}}(\tau,T,N_\tau,\tauf)/(C_F g^2)$. 
The normalization correlator in the continuum at zero flow time reads \cite{Caron-Huot:2009ncn}:
\begin{equation}
\Gnorm=\pi^2 T^4 \left[ \frac{\cos^2 \left(\pi \tau T \right)}{\sin^4\left(\pi \tau T \right) } + \frac{1}{3 \sin^2\left(\pi \tau T\right)} \right].
\end{equation}
Its lattice version at finite flow time is calculated in the same way as for the chromo-electric correlator in \cite{Altenkort:2023oms} with the difference that the observable matrix $ M_{\mu\nu}^{AB} $ here reads \cite{2022-Stendebach}
\begin{equation}
    M_{\mu\nu}^{AB}(p,\tau) = \frac{g_0^2}{18}\cos(p_0\tau)\delta^{AB}m_{\mu\nu}(p)
\end{equation}
with
\begin{equation}
\begin{gathered}
    m_{00}(p) = m_{0i}(p) = 0,\\
    m_{ij}(p) = \sum_k\hat{p}_k^2\delta_{ij} - \hat{p}_i\hat{p}_j.
\end{gathered}
\end{equation}
After tree level improvement and normalization, which can be done effectively by dividing the
the unmatched correlator by the lattice version of the normalized correlator at finite flow time, our results are shown in Fig.~\ref{fig:corr-before-cont} at two selected flow times $\sqrt{8 \tauf}/\tau = 0.25,\ 0.3$ for different temperatures. We show the results for the finest and for the coarsest lattice.
We see from the figure that the discretization effects are larger than for the chromo-electric correlator \cite{Altenkort:2023oms}. 

To perform the continuum extrapolations the correlators at different lattice spacings are first interpolated using cubic spline to the same separations $\tau T$ that are present on the finest lattice. In this way the lattice spacing effects can be seen for each $\tau T$. In Fig.~\ref{fig:cont_extra_BB} they are shown for some selected $\tau T$s at $T=195, 293$ MeV. One is the lowest and the other is the highest temperature available in this calculation. As an example we pick a fixed relative flow time $\sqrt{8 \tauf}/\tau = 0.3$. As in Ref. \cite{Altenkort:2023oms} the interpolated correlators are extrapolated to the continuum limit by fitting to the Ansatz $f(N_\tau; \tau T)=\sum_{\tau T} \left( G^\mathrm{cont}_{\tau T} - (m/(\tau T))^2 N_\tau^{-2}\right)$ in a combined way, assuming that the slope decreases
with increasing $\tau T$ \cite{Altenkort:2023oms}.
Again, it can be seen that, the lattice spacing effects are very minor, similar as in the case of the $E$-field correlators \cite{Altenkort:2023oms}. 

The obtained continuum correlators multiplied by the matching factors are independently extrapolated to the $\tauf\rightarrow 0$ limit linearly in the window $0.25\leq \sqrt{8 \tauf}/\tau \leq 0.3$. Some of the instances are shown in Fig.~\ref{fig:flow-extrap}
for $T=195$ MeV with different $\mubarir/T$ and  $\mubaruv/\muflow$. We see from the
figure that the change of $\mubaruv$ changes significantly the slope of the zero 
flow time extrapolation, while changing $\mubarir/T$ effectively amounts to a multiplicative
factor in the extrapolated value. In Fig.~\ref{fig:flow-extrap} we also show the zero
flow time extrapolation of the normalized $B$-field correlator at $T=293$ MeV. We see
that for the same choices of $\mubarir/T$ and  $\mubaruv/\muflow$ the slope of the 
zero flow time extrapolation is somewhat smaller at high temperature.

\begin{figure}[thb]
\centerline{
\includegraphics[width=0.5\textwidth]{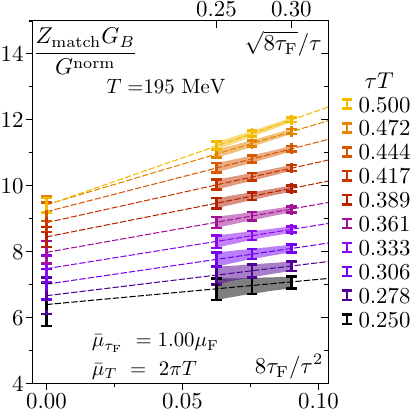}
\includegraphics[width=0.5\textwidth]{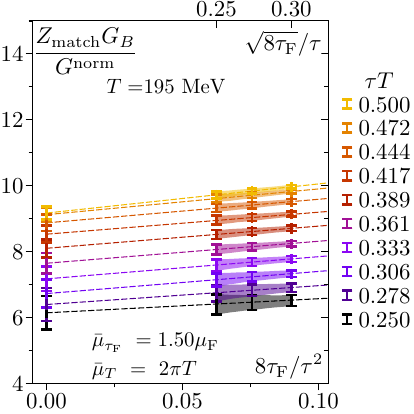}
}
\centerline{
\includegraphics[width=0.5\textwidth]{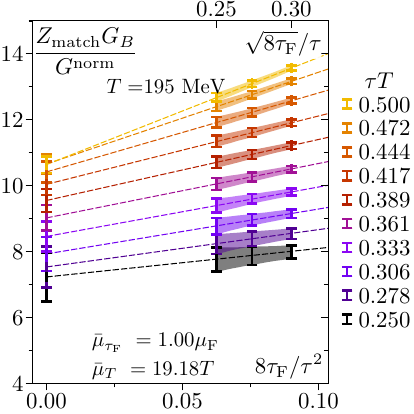}
\includegraphics[width=0.5\textwidth]{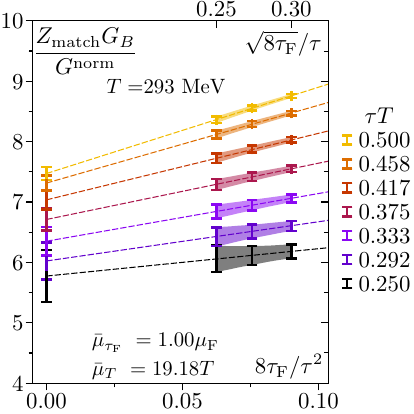}
}
\caption{$\tauf\rightarrow 0$ extrapolation of the $B$-field correlators at $T=195$ MeV with $\mubarir/T=2\pi$, $\mubaruv/\muflow=1.0$ (top left), $\mubarir/T=2\pi$, $\mubaruv/\muflow=1.5$ (top right), $\mubarir/T=19.18$, $\mubaruv/\muflow=1.0$ (bottom left) and at $T=293$ MeV with $\mubarir/T=19.18$, $\mubaruv/\muflow=1.0$ (bottom right). 
}
\label{fig:flow-extrap}
\end{figure}

\section{Spectral reconstruction}
\label{supp-mat:SPF}
As discussed in the main text, to obtain $\kappa_B$ from the $B$-field correlators
we need to model the spectral function. At high energy the spectral functions can be reliably calculated in perturbation theory.
Therefore, we could use the perturbative result to model the high energy part of the 
spectral function. The spectral function was calculated at NLO in Ref.~\cite{Banerjee:2022uge}.
Evaluating the NLO contribution numerically we found that the thermal effects are quite small,
except at very small $\omega/T$ values, where perturbation theory is not valid.
Therefore, we use the zero temperature limit of the perturbative spectral function
in $\MSBAR$ scheme at large $\omega$:
\begin{align}
\label{rhoNLO}
    \rho_B^{\mathrm{uv,LO}}(\omega,\mu) & {} = 
    \frac{\gsqms(\mu) C_{\mathrm{F}} \omega^3}{6\pi} ,
    \\
    \rho_B^{\mathrm{uv,NLO}}(\omega,\mu) & {} = %c_B^2(\mu,\mubarir)
    \frac{\gsqms(\mu) C_{\mathrm{F}} \omega^3}{6\pi}
    \left\{ 1 + \frac{\gsqms(\mu)}{(4\pi)^2} \left(
    N_c \left[ \frac 53 \ln \frac{\mu^2}{4\omega^2}
    + \frac{134}{9} - \frac{2\pi^2}{3} \right]
%    \right. \right. 
%    \nonumber \\ & {} \phantom{=} \left. \left.
    - N_f \left[ \frac 23 \ln \frac{\mu^2}{4\omega^2}
    + \frac{26}{9} \right] \right) \right\} ,
\end{align}
Here $N_c=3$, $N_f=3$ and $C_{\mathrm{F}}=(N_c^2-1)/(2N_c)$.
We note that Eq.~(\ref{rhoNLO}) contains $-2\pi^2/3$ rather than $-8\pi^2/3$ as appears in Ref.~\cite{Banerjee:2022uge} due to a mistake in the original calculation.
To obtain the spectral function for the physical scheme we need to multiply the above
expressions by $c_B(\mu,\mubarir)$ for which we use the 1-loop renormaliziation group inspired resummed expression
\begin{equation}
      c_B^2(\mu,\mubarir) =  \exp\left(\int^{\mu^2}_{\mubarir^2}\gamma_0 \gsqms(\mubar)\frac{d\mubar^2}{\mubar^2}\right).
    \label{cBNLO}
\end{equation}
After multiplying $\rho_B^{\mathrm{uv,NLO}}$ by $c_B^2(\mu,\mubarir)$ the $\MSBAR$ scale choice $\mu$ cancels at leading order, but there will be a residual $\mu$-dependence from higher order contributions. So in practice the choice of $\mu$ has an effect.
We consider two choices of $\mu$. The first choice is
$\mu^2 = 4\omega^2 + \mudr^2$, where $\mudr$ is the thermal scale of dimensionally reduced
effective field theory \cite{Kajantie:1997tt}. Our second choice is
$\mu^2 = (0.13306 \omega)^2 + \mudr^2$, which minimizes the NLO contribution for
very large $\omega$. Both choices correspond to a smooth variation of the scale
from the natural thermal scale to the frequency in the UV, thus avoiding large logs
for large $\omega$.
We also use the above choices of $\mu$ for $\rho_B^{\mathrm{uv,LO}}$. 
$\gsqms(\mu)$ is calculated in the same way as it in Eq.~(\ref{eq:Ztotal}).
To account for the missing higher order effects we multiply the UV part of the spectral
function by a constant factor, $K$.

With these choices and definitions we fit the models to our lattice data.
All models contain two fit parameters: $\kappa_B/T^3$ and $K$.
In this section we show fit results using $\mu^2=(0.13306\omega)^2+\mudr^2$, $\mubarir=19.18T$ and $\mubaruv/\muflow=1$ as an example. In Fig.~\ref{fig:fit-spf} we show the fitted spectral functions for the lowest temperature $T=195$ MeV on the left and the highest temperature $T=293$ MeV on the right. It can be seen that using LO or NLO perturbative spectral function makes only a small difference, much smaller than the overall model dependence of the spectral function.
This is because beyond the IR scale LO and NLO perturbative spectral function roughly differs by a multiplicative factor, which can be compensated by the adjustment of the $K$-factor in the fitting process, see Fig.~\ref{fig:fit-k-factor}.
%, where $K$ for NLO is always slightly smaller than that of LO.
The other two temperatures are similar. The ratio of the model correlators and the
lattice correlator is shown in 
Fig.~\ref{fig:fit-corr}, again for $T=195$ MeV and $T=293$ MeV. 
We can see that the model spectral function describes our lattice data well within errors. The obtained $\kappa_B$ from different models and scales forms a distribution, see Fig.~\ref{fig:fit-kappa}.  We calculate the weighted average and use it as our final estimate for $\kappa_B$.

\begin{figure}[thb]
\centerline{
\includegraphics[width=0.5\textwidth]{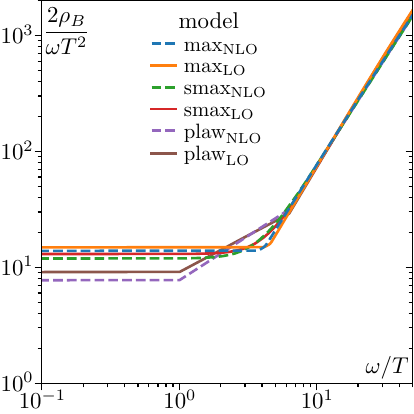}
\includegraphics[width=0.5\textwidth]{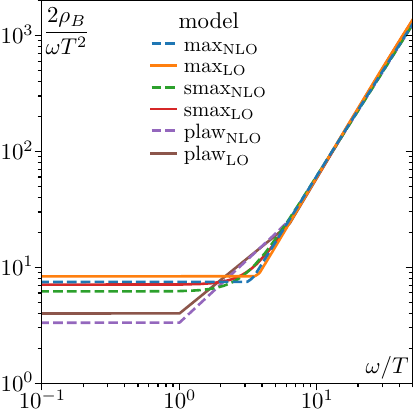}
}
\caption{The fitted spectral functions using different models for $T=195$ MeV (left) and $T=293$ MeV (right). 
}
\label{fig:fit-spf}
\end{figure}

\begin{figure}[thb]
\centerline{
\includegraphics[width=0.5\textwidth]{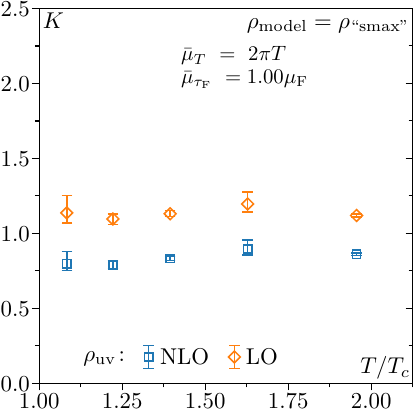}
\includegraphics[width=0.5\textwidth]{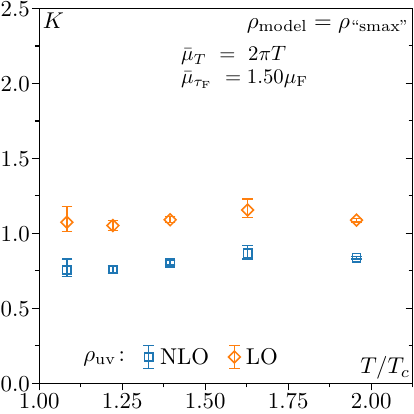}
}
\centerline{
\includegraphics[width=0.5\textwidth]{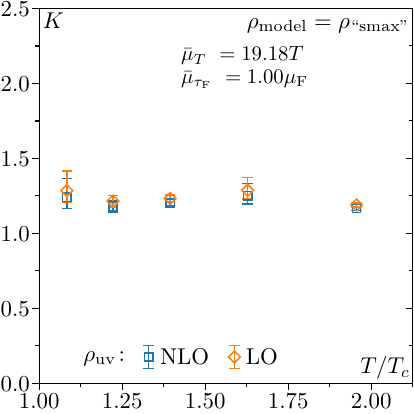}
\includegraphics[width=0.5\textwidth]{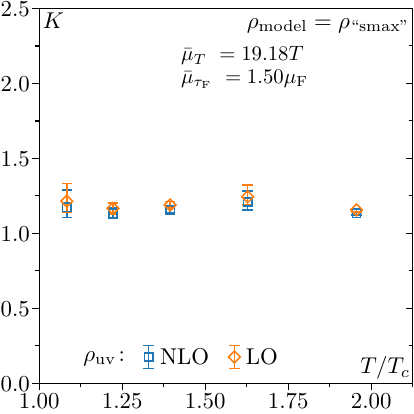}
}
\caption{The fit $K$-factor using $smax$ model at different temperatures using NLO and LO perturbative spectral function with $\mubarir/T=2\pi$, $\mubaruv/\muflow=1.0$ (top left), $\mubarir/T=2\pi$, $\mubaruv/\muflow=1.5$ (top right), $\mubarir/T=19.18$, $\mubaruv/\muflow=1.0$ (bottom left) and $\mubarir/T=19.18$, $\mubaruv/\muflow=1.5$ (bottom right). 
}
\label{fig:fit-k-factor}
\end{figure}

\begin{figure}[thb]
\centerline{
\includegraphics[width=0.5\textwidth]{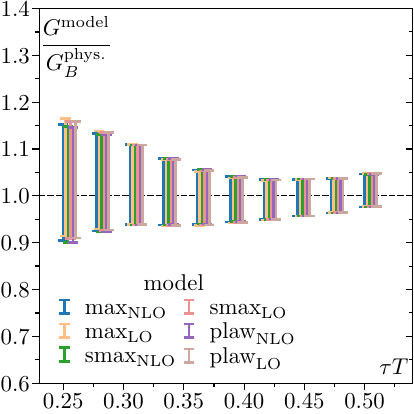}
\includegraphics[width=0.5\textwidth]{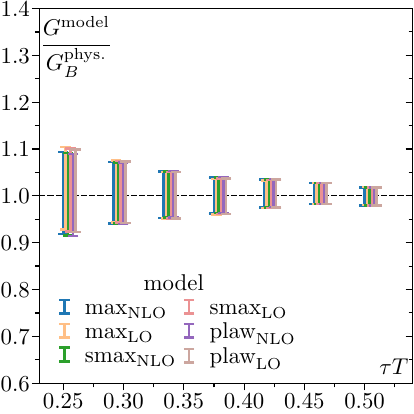}
}
\caption{The ratio of the fit correlators to the lattice correlators  at $T=195$ MeV (left) and $T=293$ MeV (right).
}
\label{fig:fit-corr}
\end{figure}

\begin{figure}[thb]
\centerline{
\includegraphics[width=0.5\textwidth]{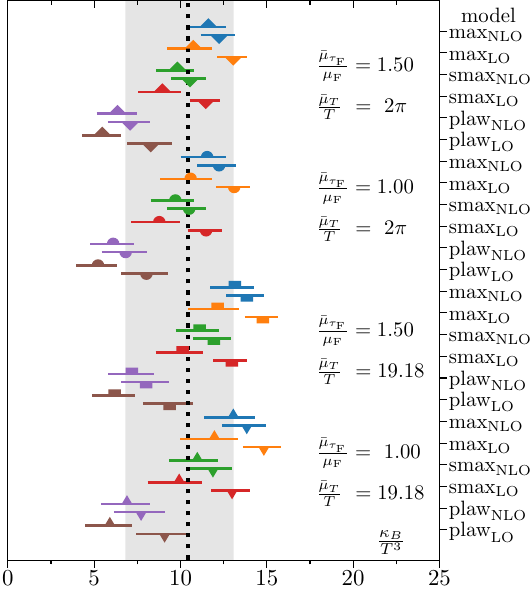}
\includegraphics[width=0.5\textwidth]{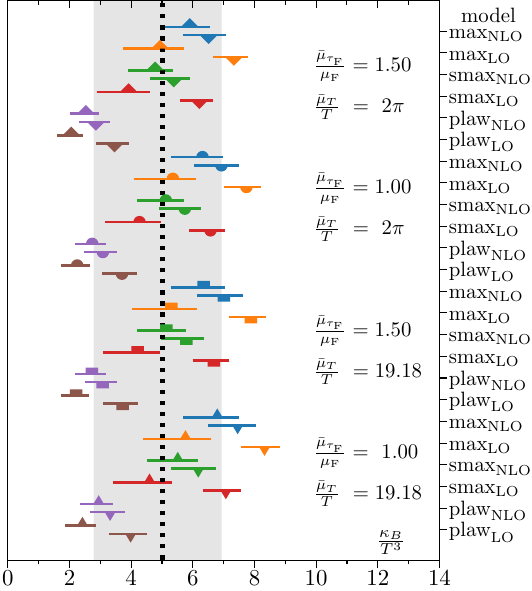}
}
\caption{The distribution of $\kappa_B$ obtained from different models and scales at $T=195$ MeV (left) and $T=293$ MeV (right). The top-half-filled points denote results using scale $\mu=\sqrt{4\omega^2+\mudr^2}$. The bottom-half-filled ones use $\mu=\sqrt{(0.13306\omega)^2+\mudr^2}$. The gray band denotes the weighted average regarded as confidence interval drawn from the distribution.
}
\label{fig:fit-kappa}
\end{figure}

\section{Comparison of $\kappa_E$ and $\kappa_B$}
\label{sec:compare-kappa}
In Fig.~\ref{fig:compare-kappa} we compare $\kappa_E$ and $\kappa_B$ obtained in quenched lattice QCD and $2+1$-flavor lattice QCD. The quenched $\kappa_E$ are taken from 
\cite{Altenkort:2020fgs,Brambilla:2020siz,Banerjee:2022gen} calculated using gradient flow or multilevel algorithm. The full QCD $\kappa_E$ is taken from Ref.~\cite{Altenkort:2023oms}. For quenched $\kappa_B$ we select two estimates, one from multilevel algorithm \cite{Banerjee:2022uge} and the other from gradient flow at finite $\tauf$ \cite{Brambilla:2022xbd}. We present the figures in terms of both $T/T_c$ and $T$. One can see that $\kappa_E$ and $\kappa_B$ are of similar size, in both quenched limit and full QCD, when shown in $T/T_c$. The full QCD results are much larger at small $T/T_c$. When presented in $T$, both $\kappa_E/T^3$ and $\kappa_B/T^3$ follow a decreasing pattern, connecting smoothly to the quenched results at high temperature.

\begin{figure}[tbh]
\centerline{
\includegraphics[width=0.5\textwidth]{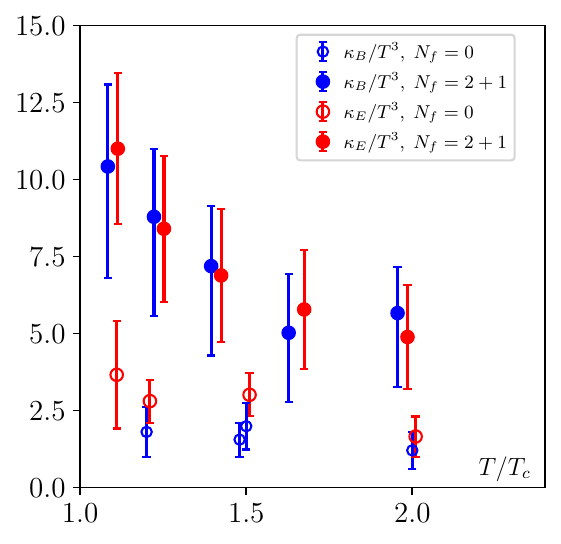}
\includegraphics[width=0.5\textwidth]{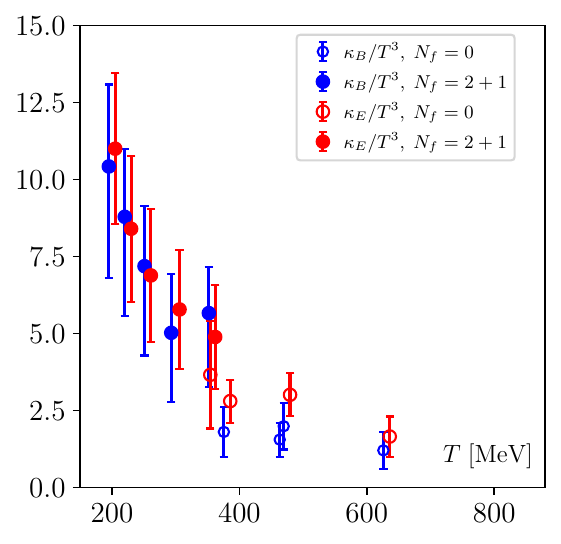}}
\caption{Comparison of $\kappa_{E,B}$ from lattice calculations in terms of $T/T_c$ (left) and $T$ (right). The quenched $\kappa_E$ are taken from \cite{Altenkort:2020fgs,Brambilla:2020siz,Banerjee:2022gen} and the 2+1 flavor $\kappa_E$ is taken from \cite{Altenkort:2023oms}. The quenched $\kappa_B$ are taken from \cite{Banerjee:2022uge,Brambilla:2022xbd}. The $\kappa_E$ points are slightly shift horizontally for better visibility.}
\label{fig:compare-kappa}
\end{figure}

\section{Estimation of $\langle v^2\rangle$ and $\langle p^2\rangle$}
\label{sec:mass-dep}
To obtain $\kappa$ and $D_s$ we need to evaluate the averaged velocity squared, $\langle v^2\rangle$,
and the averaged momentum squared, $\langle p^2\rangle$ of the heavy quark. This can be done
assuming that the heavy quarks are well defined quasi particles in the hot medium with a potentially
temperature dependent mass, $M$ \cite{Petreczky:2008px}. Thus we need to define the masses of the charm and bottom quarks.

To obtain the charm quark mass, we fit the continuum-extrapolated lattice QCD results of the charm quark number susceptibility~\cite{Bellwied:2015lba} to the quasi particle model~\cite{Petreczky:2008px}
\begin{equation}
\frac{\chi_2^\text{charm}}{T^2}=\frac{4N_c}{(2\pi T)^3}\int d^3p\ e^{-E_p/T},
\end{equation}
where $E_p^2=M^2+p^2$. Here we use Boltzmann approximation because the heavy quark
masses are significantly larger than the temperature.
The lattice data of the susceptibility and the fit results are shown in the left panel of Fig.~\ref{fig:mass-fit} and the extracted charm quark mass is shown in the right panel. It can be seen that the quasi particle model describes the lattice data rather well. 
We also see that the temperature variation of the charm quark mass between $T=190$ MeV
and $T=300$ MeV is 309 MeV.
Then the mass is interpolated to the desired temperatures, listed in Table ~\ref{tab:bottom-charm-mass}. In the table we indicate the uncertainty of the charm
masses due to the uncertainties in $\chi_2^{charm}$.
\begin{figure}[tbh]
\centerline{
\includegraphics[width=0.5\textwidth]{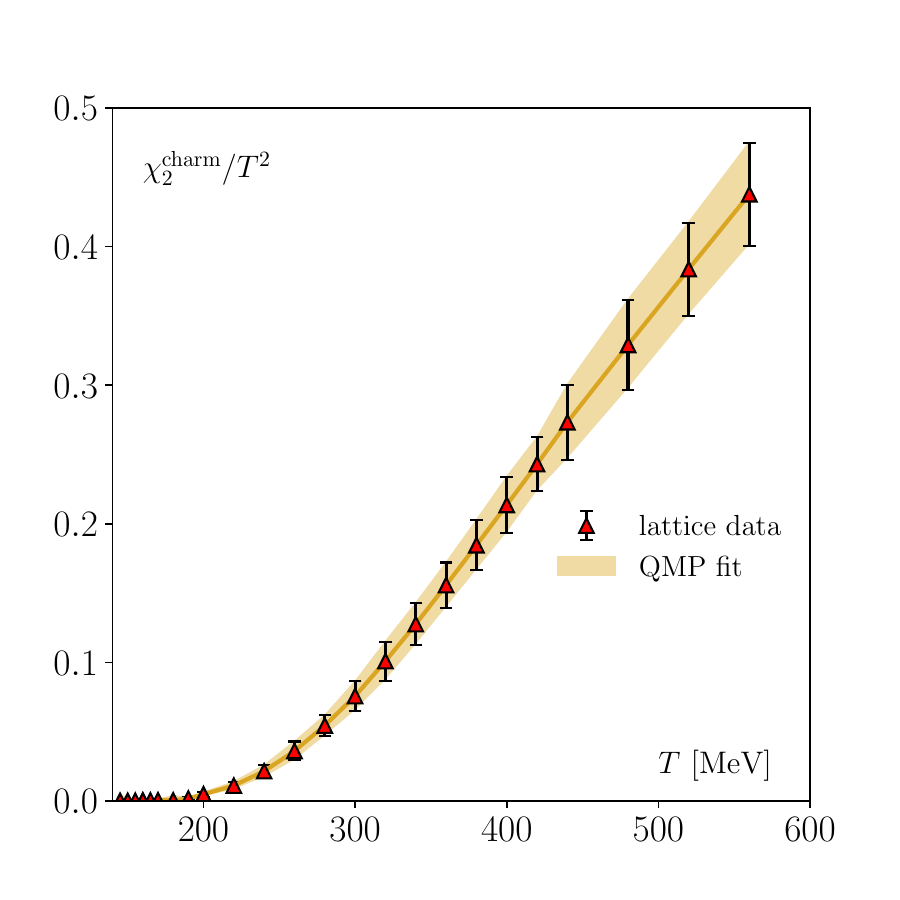}
\includegraphics[width=0.5\textwidth]{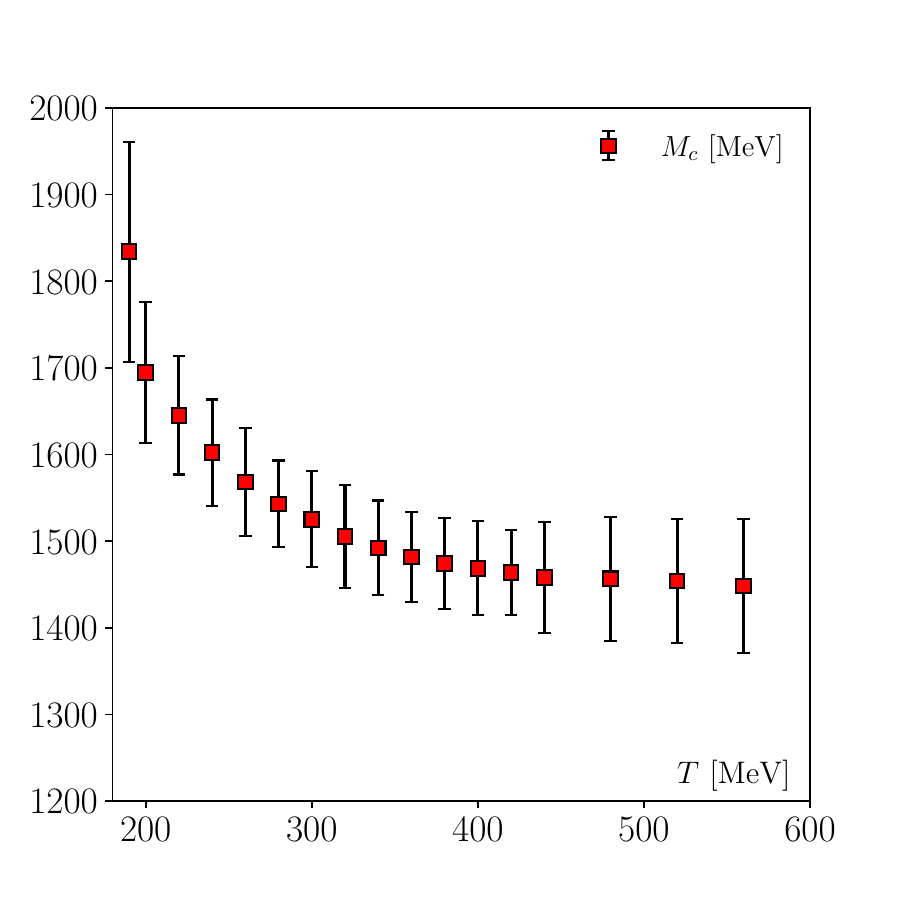}}
\caption{Left: lattice results of the charm quark susceptibility \cite{Bellwied:2015lba} and the fit to quasi particle model. Right: the extracted charm quark mass from the fits.}
\label{fig:mass-fit}
\end{figure}
With the extracted mass, $\langle v^2\rangle$ and $\langle p^2\rangle$ can be calculated again in quasi-particle model~\cite{Petreczky:2008px}:
\begin{equation}
\begin{split}
 \langle v^2\rangle &=\left(\int d^3p\ \frac{p^2}{E^2_p}e^{-E_p/T}\right)/\left(\int d^3p\ e^{-E_p/T}\right), \\
 \langle p^2\rangle &=\left(\int d^3p\ p^2e^{-E_p/T}\right)/\left(\int d^3p\ e^{-E_p/T}\right).
 \end{split}
\end{equation}
In Table~\ref{tab:bottom-charm-mass} we show  $\langle v^2\rangle$, $\langle p^2\rangle/(3MT)$ and 
$D_s$ for charm quarks. Note that the uncertainties in these quantities due to the uncertainty
of $M_c$ are small.

We consider two values 4.5 GeV and 4.8 GeV for the bottom quark. 
The difference between these two values of the bottom quark mass should cover possible
thermal effects, see above. In Table~\ref{tab:bottom-charm-mass} we also show  $\langle v^2\rangle$, $\langle p^2\rangle/(3MT)$ and 
$D_s$ for bottom quarks. Again the effects of the choice of $M_b$ are small.

\begin{table}[htb]
\centering
\begin{tabular}{|c|cccc|cccc|}
\hline
$T$ [MeV] & $M_b/T$ & $\langle v^2\rangle$ & $\langle p^2\rangle/(3MT)$ & $2\pi TD_s$ & $M_c$ [GeV] & $\langle v^2\rangle$ & $\langle p^2\rangle/(3MT)$ & $2\pi TD_s$\\
\hline
195 & 23.08 & 0.117 & 1.112 &	1.242(267) & 1.74(12) & 0.264(15) & 1.303(23) & 1.338(279)(12)\\
    & 24.62 & 0.111 & 1.105 &	1.240(267) & & & & \\
\hline
220 & 20.45 & 0.131 & 1.127 &	1.671(451) & 1.65(7) & 0.302(9) & 1.364(16) & 1.817(468)(11)\\
    & 21.82 & 0.124 & 1.118 &	1.666(450) & & & & \\
\hline 
251 & 17.93 & 0.147 & 1.145 &	2.093(627) & 1.58(5) & 0.342(8) & 1.438(16) & 2.335(671)(14)\\
    & 19.12 & 0.139 & 1.136 &	2.088(626) & & & & \\
\hline
293 & 15.35 & 0.168 & 1.170 &	2.598(835) & 1.53(5) & 0.390(9) & 1.536(19) & 3.026(938)(22)\\
    & 16.38 & 0.159 & 1.159 &	2.584(831) & & & & \\
    \hline
352 & 12.78 & 0.197 & 1.206 &	3.004(947) & 1.49(5) & 0.448(9) & 1.677(23) & 3.581(1108)(27)\\
    & 13.64 & 0.186 & 1.193 &	2.993(946) & & & & \\
    \hline     
\end{tabular}
\caption{The dependence of $\langle v^2\rangle$, $\langle p^2\rangle/(3MT)$ and $2\pi TD_s$ on the quark mass. The four columns in the middle are for the bottom quark. The uncertainties of $2\pi TD_s$ are from $\kappa$ solely. The first row at each temperature is for $m_b$=4.5 GeV while the second row for 4.8 GeV. The four columns on the right are for the charm quark. Note that the uncertainties in the first brackets of the last column inherit from $\kappa$ and the second brackets from the uncertainties in the charm quark mass.}
\label{tab:bottom-charm-mass}
\end{table}

\end{widetext}
\end{document}